\documentclass[acmsmall]{acmart}

\usepackage{booktabs}
\usepackage{colortbl}
\usepackage{enumitem}
\usepackage{makecell}
\usepackage{rotating}
\usepackage{multirow}
\usepackage{tikz}
\usepackage{url}
\usepackage{fancybox}
\usepackage{calc}
\usepackage{tabularx}
\usepackage{bbm}
\usepackage{listings}
\usepackage{xcolor}
\usepackage{tcolorbox}
\usepackage{pifont}

\usetikzlibrary{shapes.misc}
\usetikzlibrary{shadows}

\definecolor{lightgray}{gray}{0.9}
\newcolumntype{Y}{>{\raggedright\arraybackslash}X}

\lstdefinestyle{minimal}{
  basicstyle=\ttfamily\scriptsize,
  aboveskip=0pt,
  belowskip=0pt,
  frame=none,
  breaklines=true,
  postbreak=\mbox{\textcolor{red}{$\hookrightarrow$}\space},
}

\lstdefinestyle{greenHighlight}{
  basicstyle=\ttfamily\scriptsize,
  moredelim=**[is][\color{teal}]{@}{@},
}

\lstdefinestyle{redHighlight}{
  basicstyle=\ttfamily\scriptsize,
  moredelim=**[is][\color{purple}]{@}{@},
}

\tcbset{arc=2pt, boxrule=\heavyrulewidth, colback=gray!1, colframe=black}

\newcommand{\codeblock}[1]{%
  \begingroup
  \ttfamily\scriptsize
  \begin{minipage}[t]{\linewidth}%
  \raggedright #1%
  \end{minipage}%
  \endgroup
}

\newcommand{\RowGap}{\addlinespace[6pt]}

\newcommand{\rqz}{How good are LLMs in generating Cypher queries for knowledge graphs?}
\newcommand{\rqi}{How effective is our approach in answering software repository-related questions?}
\newcommand{\rqii}{What are the limitations of our approach in accurately answering software repository-related questions?}
\newcommand{\rqiii}{Can chain-of-thought prompting improve the effectiveness of our approach in answering software repository-related questions?}
\newcommand{\rqiv}{How do users perceive the usefulness of our approach in assisting them to answer repository-related questions?}

\newcommand{\cmark}{\ding{51}}
\newcommand{\xmark}{\ding{55}}

\begin{document}

\title{Synergizing LLMs and Knowledge Graphs: A Novel Approach to Software Repository-Related Question Answering}

\author{Samuel Abedu}
\email{samuel.abedu@mail.concordia.ca}
\orcid{0009-0000-0472-4514}
\affiliation{%
  \department[0]{Data-driven Analysis of Software (DAS) Lab}
  \department[1]{Department of Computer Science \& Software Engineering}
  \institution{Concordia University}
  \city{Montreal}
  \state{QC}
  \country{Canada}}

\author{SayedHassan Khatoonabadi}
\email{sayedhassan.khatoonabadi@concordia.ca}
\orcid{0000-0003-0615-9242}
\affiliation{%
  \department[0]{Data-driven Analysis of Software (DAS) Lab}
  \department[1]{Department of Computer Science \& Software Engineering}
  \institution{Concordia University}
  \city{Montreal}
  \state{QC}
  \country{Canada}}

\author{Emad Shihab}
\email{emad.shihab@concordia.ca}
\orcid{0000-0003-1285-9878}
\affiliation{%
  \department[0]{Data-driven Analysis of Software (DAS) Lab}
  \department[1]{Department of Computer Science \& Software Engineering}
  \institution{Concordia University}
  \city{Montreal}
  \state{QC}
  \country{Canada}}

\renewcommand{\shortauthors}{Abedu et al.}

\begin{abstract}
Software repositories contain valuable information for understanding the development process. However, extracting insights from repository data is time-consuming and requires technical expertise. While software engineering chatbots support natural language interactions with repositories, chatbots struggle to understand questions beyond their trained intents and to accurately retrieve the relevant data. This study aims to improve the accuracy of LLM-based chatbots in answering repository-related questions by augmenting them with knowledge graphs. We use a two-step approach: constructing a knowledge graph from repository data, and synergizing the knowledge graph with an LLM to handle natural language questions and answers. We curated 150 questions of varying complexity and evaluated the approach on five popular open-source projects. Our initial results revealed the limitations of the approach, with most errors due to the reasoning ability of the LLM. We therefore applied few-shot chain-of-thought prompting, which improved accuracy to 84\%. We also compared against baselines (MSRBot and GPT-4o-search-preview), and our approach performed significantly better. In a task-based user study with 20 participants, users completed more tasks correctly and in less time with our approach, and they reported that it was useful. Our findings demonstrate that LLMs and knowledge graphs are a viable solution for making repository data accessible.
\end{abstract}



\maketitle

\section{Introduction}
\label{sec:introduction}
Software repositories are rich sources of information essential to the software development process. This includes data on source code, documentation, issue tracking data, and commit histories~\cite{vidoniSystematicProcessMining2022}. Analyzing this data can provide valuable insights about a project, such as developer activities and project evolution~\cite{hassanRoadAheadMining2008}. For instance, \citet{begelAnalyzeThis1452014} and \citet{sharmaWhatDevelopersWant2017} presented questions that software practitioners are interested in answering regarding their projects. Question answering is a core part of day-to-day development: prior studies show that developers continuously seek answers about project state, processes, and code to make progress, underscoring the need for tools that surface precise, timely answers~\cite{begelAnalyzeThis1452014,ko_information_2007}. Answering some of these questions requires mining and analyzing repository data. However, accessing and extracting meaningful insights from repositories is time-consuming and requires technical expertise~\cite{abdellatifMSRBotUsingBots2020,banerjeeCostMiningVery2015}. For example, in a StackOverflow post~\cite{izkerosHowCanCalculate2022}, a user seeking to calculate the number of lines changed since the last commit in a Git repository found that the solution required using specific Git commands like \texttt{git diff --shortstat}, which can be challenging for non-technical stakeholders. The technical knowledge and the time spent on such a task can be a barrier to software practitioners.

Prior studies have attempted to address this challenge by developing software engineering chatbots that provide intuitive, natural language interfaces to software repositories~\cite{abdellatifMSRBotUsingBots2020,abeduLLMBasedChatbotsMining2024}. However, a key challenge in software engineering chatbot development lies in natural language understanding (NLU), as the chatbot should accurately interpret user questions and map them to appropriate data retrieval actions~\cite{abdellatifChallengesChatbotDevelopment2020}. Additionally, the NLU approach to chatbot development fails when the NLU model is not trained on the intent of the user's question. Each intent is typically mapped to a predefined action within the chatbot's framework~\cite{rameshSurveyDesignTechniques2017,adamopoulouOverviewChatbotTechnology2020}. However, it is often impractical to define actions for every possible intent, especially as user requirements evolve, limiting the chatbot's functionality and adaptability. Large language models (LLMs) have demonstrated remarkable capabilities in understanding natural language and identifying the intents of input texts~\cite{shahUsingLargeLanguage2024}. Nonetheless, leveraging LLMs to build chatbots for repository question answering using the naive retrieval-augmented generation (RAG) approach has proved challenging. \citet{abeduLLMBasedChatbotsMining2024} reported that LLM-based RAG chatbots failed to retrieve accurate data to answer repository-related questions 83.3\% of the time.

Knowledge graphs have the potential to enhance LLMs with external data to generate contextually relevant responses~\cite {panUnifyingLargeLanguage2024,abu-rasheedKnowledgeGraphsContext2024}. Knowledge graphs are structured representations that model entities and their relationships, enabling enhanced semantic and structural understanding and reasoning~\cite{jiSurveyKnowledgeGraphs2022,chenReviewKnowledgeReasoning2020,xieRepresentationLearningKnowledge2016}. Due to the structured nature of software repositories, prior studies have modelled software repositories as knowledge graphs to solve various problems in the software engineering domain~\cite{malikGraphanonymizationSoftwareAnalytics2024, zhaoKnowledgeGraphingGit2019,maHowUnderstandWhole2024}. Motivated by this, we aim to improve the accuracy of LLM-based chatbots in answering repository-related questions by synergizing LLMs with knowledge graphs.
In this study, similar to prior studies, we limit the knowledge graph modelling to Git metadata~\cite{malikGraphanonymizationSoftwareAnalytics2024} and focus on answering repository-related questions that are limited to Git metadata, similar to~\cite{abdellatifMSRBotUsingBots2020,abeduLLMBasedChatbotsMining2024}.

Our approach is comprised of two key steps: (1) data ingestion and (2) interaction. The data ingestion step has one component, the Knowledge Graph Constructor, which collects and models repository data to construct a knowledge graph. The interaction step consists of three components: the Query Generator, which translates natural language questions into graph queries using an LLM; the Query Executor,  which extracts and runs the graph queries against the knowledge graph to retrieve relevant information; and the Response Generator, which generates an answer to the user's question based on the retrieved data using an LLM.

Our approach relies on the ability of LLMs to generate correct graph queries from natural language. Therefore, we conducted an exploratory analysis to assess and identify the most efficient state-of-the-art LLM for graph query generation from natural language input. We evaluated several LLMs, including GPT-4o~\cite{openaiHelloGPT4o2024}, Llama3~\cite{metaMetallamaMetaLlama38BHugging2024}, and Claude3.5~\cite{anthropicIntroducingClaude352024}, on their ability to generate accurate graph queries which we operationalize using Cypher~\cite{francisCypherEvolvingQuery2018} from natural language text and found GPT-4o the most accurate in our context. 
We integrated GPT-4o with the knowledge graph–based approach and evaluated it on five popular open-source repositories (AutoGPT, Bootstrap, Ohmyzsh, React, Vue) using 150 questions curated from~\citet{abdellatifMSRBotUsingBots2020}. We first selected a subset of 20 questions spanning the intents and difficulty levels to run the initial evaluation and identify failure modes; we then ran the best configuration on the full set of 150 questions. Questions were grouped into three difficulty levels based on the number of relationships in the knowledge graph needed to retrieve the data. Each question was executed five times with a majority vote for correctness. We compared against MSRBot and GPT-4o-search-preview as baselines and conducted a task-based user study with 20 participants. Specifically, our evaluation aims to answer the following research questions:

\begin{itemize}
    \item[\textbf{RQ1:}] \textbf{\rqi} We find that synergizing LLMs with knowledge graphs correctly answered repository-related questions 65\% of the time. The approach performs well when answering simple questions but struggles with complex questions that require two or more relationships from the knowledge graph.
    \item[\textbf{RQ2:}] \textbf{\rqii} We find the reasoning of LLM during query generation is the most prevalent limitation affecting LLMs when enhanced with knowledge graphs. It hinders the ability of the LLM to accurately interpret and utilize the right nodes and relationships within the knowledge graph, leading to incorrect relationship modeling, faulty arithmetic logic, misapplied attribute filtering, and misapplied date formatting. Other limitations include the LLM making wrong assumptions and hallucination.
    \item[\textbf{RQ3:}] \textbf{\rqiii} We find that the chain-of-thought prompting approach answered the repository-related question 84\% of the time, up from an initial 65\% without chain-of-thought. Specifically, the accuracy of the complex questions requiring two or more relationships increased from 50\% to 90\%. This implies that chain-of-thought can help answer complex questions requiring multiple relationships. 
    \item[\textbf{RQ4:}] \textbf{\rqiv} We find that the study participants completed more tasks correctly and in less time with the chatbot than with their usual methods, and they perceived the chatbot as useful and time-saving.
\end{itemize}

Our findings demonstrate that the synergy of LLMs, knowledge graphs, and chain-of-thought prompting can be effective in answering repository-related questions. In summary, we make the following contributions in this paper: 
\begin{itemize}
    \item We provide empirical evidence demonstrating the capabilities of LLMs in generating Cypher queries for querying software repository data stored/represented in knowledge graphs.
    \item We discuss the limitations of augmenting LLMs with software repository data stored/represented in knowledge graphs.
    \item To the best of our knowledge, this is the first software repository question-answering approach based on knowledge graphs.
    \item We share the dataset and scripts for reproducibility and advancing the field at~\cite{abeduSabeduKnowledge_graph_llm_synergyReplication2024}.
\end{itemize}

\noindent\textbf{Paper organization.} The rest of the paper is structured as follows: We begin by explaining the concepts in this paper with the related works in Section~\ref{sec:background}. We present our approach in Section~\ref{sec:approach} and the evaluation set-up in Section~\ref{sec:eval_setup}. We present the results of our research questions in Section~\ref{sec:results} and discuss the findings in Section~\ref{sec:discussions}. We outline the threats that can affect the validity of our results in Section~\ref{sec:threats} and conclude the paper in Section~\ref{sec:conclusion}.

\section{Background \& Related Works}
\label{sec:background}
In this section, we provide an overview of the key concepts that form the foundation of our study. We discuss software repositories and their significance, software engineering chatbots, and knowledge graphs.

\subsection{Software Repositories}
Software repositories contain data that track the development process of a project~\cite{hassanRoadAheadMining2008}. Platforms like GitHub and Jira provide version control systems that facilitate collaboration among developers, track changes over time, and support issue tracking and project management. Software repositories contain a wealth of information, including details about commits, pull requests, issues, and developer activities.

Prior studies have analyzed repository data to investigate and understand various development processes. For instance, \citet{dilharaUnderstandingSoftware2Study2021} conducted a large-scale analysis of commit data on GitHub to understand the evolution of machine learning library usage in open-source projects. \citet{hataGitHubDiscussionsExploratory2021} conducted a mixed-methods study to understand how developers use GitHub's Discussions feature by analyzing early adopters. \citet{khatoonabadiPredictingFirstResponse2024} utilized pull request data from 20 open-source projects from GitHub to develop a machine learning approach to predict the first response latency of both maintainers and contributors during the pull request review.

Accessing and interpreting this information is crucial for various stakeholders, including developers, project managers, and non-technical team members. However, analyzing and extracting meaningful insights from these data can be challenging without specialized knowledge and also time-consuming~\cite{banerjeeCostMiningVery2015,abdellatifMSRBotUsingBots2020}.

\subsection{Software Engineering Chatbots}
Chatbots are conversational assistants designed to assist with specific tasks by interacting with users through natural language~\cite{rameshSurveyDesignTechniques2017}. They aim to facilitate access to information, automate routine tasks, and support collaboration among team members~\cite{abdellatifComparisonNaturalLanguage2022}. 
Evidence from recent studies underscores that question-answering has become central to developer workflows: controlled and field studies report sizable productivity gains from conversational assistants, while surveys show widespread but cautious adoption of AI assistants for code search, explanation, and guidance~\cite{khojah_beyond_2024,peng_impact_2023,stackoverflow_ai_2024}. Chatbots are increasingly becoming popular in the software engineering domain to accomplish specific software engineering tasks. For instance, \citet{abdellatifMSRBotUsingBots2020} proposed MSRBot, using a bot layered on top of software repositories to automate and simplify the extraction of useful information from the repository. \citet{bradleyContextawareConversationalDeveloper2018} proposed Devy, a Conversational Developer Assistant that enables developers to focus on high-level tasks by reducing the need for manual low-level commands across various tools. \citet{dominicConversationalBotNewcomers2020} proposed a conversational bot to support newcomers in onboarding to open-source projects by recommending suitable projects, resources, and mentors. \citet{okanovicCanChatbotSupport2020} proposed PerformoBot, a chatbot that guides developers through configuring and executing load tests via natural language conversations. Also, \citet{abeduLLMBasedChatbotsMining2024} developed an LLM-based chatbot to answer questions related to software repositories. Their LLM-based chatbot, which used the RAG approach, failed to retrieve the relevant data needed to answer questions in their evaluation questions most of the time.
Beyond these task-specific bots, recent code-aware assistants such as GitHub Copilot Chat, CodeT5+, and StarCoder bring conversational capabilities into the IDE, supporting code generation, explanation, and navigation; however, they primarily target source-level reasoning rather than repository metadata Q\&A across issues and commits~\cite{zhuo_bigcodebench_2025,wangCodeT5OpenCode2023}.

The increasing application of chatbots in software engineering and LLMs in chatbots like ChatGPT and BARD motivates our work to improve the accuracy of LLMs in software engineering chatbots. To the best of our knowledge, this is the first software repository question-answering approach based on knowledge graphs and LLMs.

\citet{abdellatifMSRBotUsingBots2020} and \citet{abeduLLMBasedChatbotsMining2024} are closest to our work. Like MSRBot~\cite{abdellatifMSRBotUsingBots2020}, we follow a similar interaction pattern, where a user asks a question; the approach queries a repository’s data and then generates an answer for the user. Also, we evaluate our approach using the evaluation dataset from~\citet{abdellatifMSRBotUsingBots2020}. However, unlike~\citet{abdellatifMSRBotUsingBots2020}’s fixed intent/entity pipeline, our work uses an LLM generator with a repository knowledge graph, enabling support for more user questions with variable intents. \citet{abeduLLMBasedChatbotsMining2024} showed that RAG-based LLM chatbots often fail because they retrieve irrelevant context; we address this by grounding the LLM in a structured knowledge graph that supplies precise repository facts rather than unstructured snippets, reducing retrieval mismatch and improving answer reliability~\cite{panUnifyingLargeLanguage2024,abu-rasheedKnowledgeGraphsContext2024}.
While Copilot-style systems allow users to interact with their repository at the code level, our focus is on allowing users to interact at the repository-metadata level.

\subsection{Knowledge Graphs and Large Language Models}
Knowledge graphs are structured representations of information that model entities (nodes) and the relationships (edges) between them~\cite{hoganKnowledgeGraphs2021}. They effectively organize and represent knowledge as triple facts (\textit{head entity}, \texttt{relationship}, \textit{tail entity}), allowing it to be efficiently utilized in advanced applications~\cite{chenReviewKnowledgeReasoning2020,xieRepresentationLearningKnowledge2016}. Popularized by Google's introduction in 2012~\cite{singhalIntroducingKnowledgeGraph2012}, knowledge graphs have been widely used in domains such as the semantic web, natural language processing, and recommendation systems~\cite{jiSurveyKnowledgeGraphs2022}.

In the software engineering domain, prior studies have represented software repositories as knowledge graphs. For instance, \citet{zhaoKnowledgeGraphingGit2019} proposed GitGraph, a prototype tool that automatically constructs knowledge graphs from Git repositories to help developers and project managers comprehend software projects. \citet{malikGraphanonymizationSoftwareAnalytics2024} introduced a method for representing software repositories as graphs to preserve the context between different features during anonymization for data sharing in software analytics. Additionally, \citet{maHowUnderstandWhole2024} developed RepoUnderstander, a method that condenses critical information from entire software repositories into a repository knowledge graph to guide agents in comprehensively understanding the repositories.

By structuring repository data into a knowledge graph, it becomes possible to perform complex queries and infer new knowledge through graph traversal and pattern matching. Query languages like Cypher, used with graph databases such as Neo4j, Redis graphs, and MemGraph, enable querying of knowledge graphs using declarative language~\cite{francisCypherEvolvingQuery2018}.

LLMs are effective at natural language but can hallucinate, lack provenance, and provide out-of-date information. Knowledge graphs provide explicit and verifiable facts with typed relations and support multi-hop querying. Integrating the two technologies improves LLM outputs and accuracy.
Prior studies on this combination outline these benefits. For instance, \citet{li_graphotter_2024} converts tables into graphs and guides step-by-step reasoning, which filters noise and yields higher QA accuracy than text-based baselines. \citet{xu_harnessing_2025} retrieve entities, relations, and subgraphs and align them before prompting, which improves answer accuracy and logical form over text retrieval. \citet{lavrinovics_knowledge_2025} review knowledge graph-based strategies that reduce hallucinations during pretraining, inference, and post-generation, and they call for stronger evaluation. \citet{hogan_large_2025} show that systems that delegate multi-hop facts to knowledge graphs, freshness to search, and fluency to LLMs deliver more reliable answers. \citet{sequeda_knowledge_2025} argue that knowledge graphs enable trust and provenance in enterprise QA, and they report better auditability and governance when answers are checked against knowledge graph facts.

For repository question answering, combining LLMs with a knowledge graph is a viable approach because a repository knowledge graph provides verifiable facts that ground the LLMs’ responses and improve answer accuracy. To the best of our knowledge, this is the first study to address repository question answering using an LLM enhanced with a knowledge graph.

\subsection{Prompt Engineering for LLMs in Software Engineering}
Large language models are widely used for software engineering tasks such as code generation, summarization, and commenting. Prompt engineering guides these models with task-specific instructions without changing model weights. A recent systematic review by~\citet{hou_large_2024} lists the common prompt engineering techniques used in recent software engineering studies. These include zero-shot, few-shot, chain-of-thought (CoT), automatic prompt engineering (APE), prompt-based continuous prompting (PromptCS), Chain-of-Code (CoC), modular-of-thought (MoT), and structured chain-of-thought (SCoT), and reports the broad use of few-shot and CoT prompting~\cite{hou_large_2024}.

\citet{geng_large_2024} investigate multi-intent comment generation and report that providing ten or more in-context examples in the prompt (few-shot learning) enables diverse comments and outperforms a supervised baseline. \citet{xu_unilog_2024} propose UniLog for automatic logging and show that five demonstration examples in the prompt (few-shot learning), with example selection and ordering, improve insertion quality without model tuning. \citet{wu_how_2023} study security repair and use prompts that mark buggy and fixed regions (e.g., “BUG:”/“FIXED:”), finding that success is limited and mostly on simple cases.

\citet{shin_prompt_2025} compares GPT-4 under basic, in-context, and task-specific prompts to fine-tuned models across summarization, generation, and translation, and reports that prompt-engineered GPT-4 is on par with fine-tuned models and offers ease of use, especially with a natural language interface.
These studies show that prompt engineering can improve results of LLM applications on SE tasks, especially with few-shot and CoT prompting~\cite{hou_large_2024,shin_prompt_2025}.

\section{Approach}
\label{sec:approach}
\begin{figure}
\centering
\includegraphics[width=.95\columnwidth]{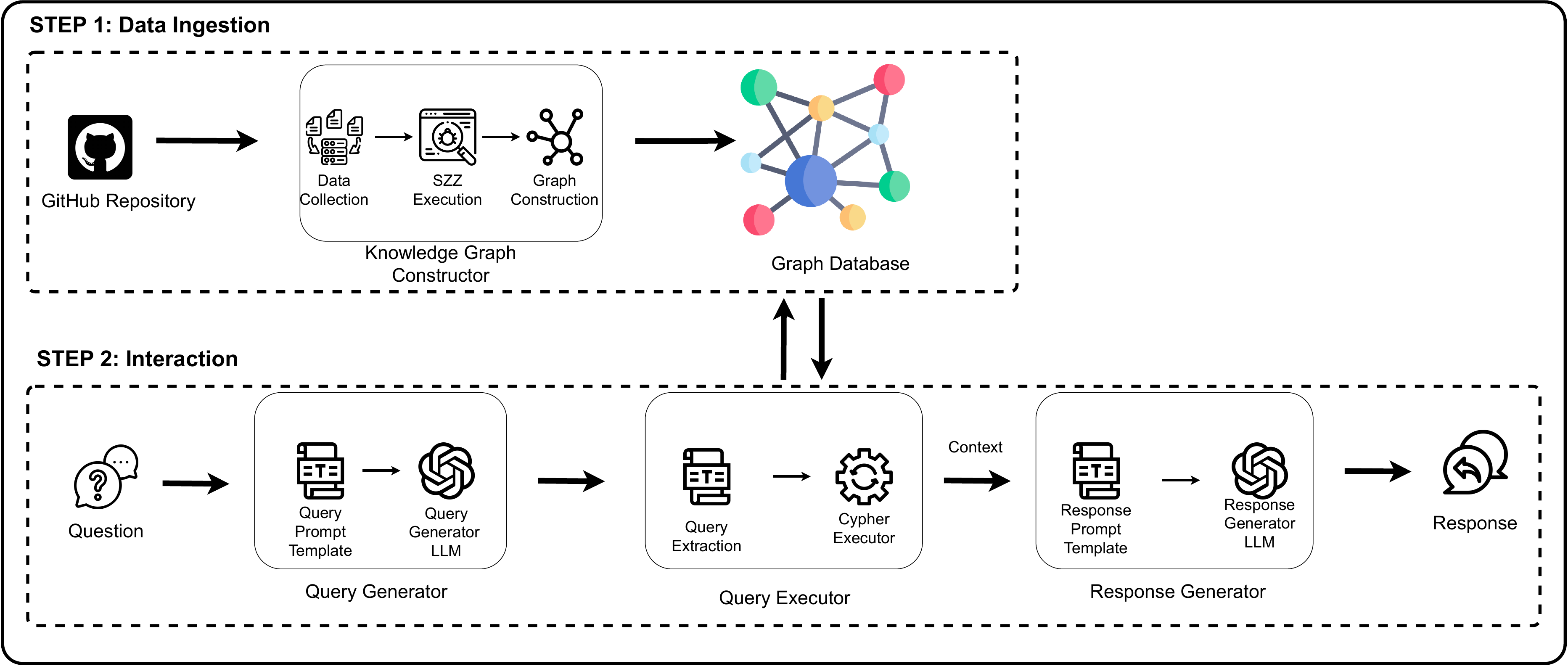}
\caption{Overview of our approach in answering software repository-related questions by synergizing LLMs and knowledge graphs.}
\label{fig:approach_overview}
\end{figure}

Figure \ref{fig:approach_overview} provides an overview of our approach to answering repository-related questions. Our approach consists of four key components organized into two steps: (1) data ingestion and (2) interaction. In the data ingestion step, the \textbf{Knowledge Graph Constructor} component collects repository data and models it as a knowledge graph. During the interaction step, the \textbf{Query Generator} component takes the user’s natural language question as input and generates a graph query using an LLM to retrieve the relevant data required to answer the question. The \textbf{Query Executor} component then takes the generated query from the Query Generator component and executes it. It returns the results of the query, which are used by the \textbf{Response Generator} component as context to generate a natural language response to the user's question using an LLM. In this section, we describe each component of our approach in detail, using the question \textit{``How many people have contributed to the code?''} as our running example.

\subsection{Knowledge Graph Constructor}
\label{sec:kg_constructor}
The Knowledge Graph Constructor component aims to connect the entities in the software repository to form the repository knowledge graph. Using a knowledge graph allows us to model the complex relationships between the repository entities, facilitating analysis and inference of the repository data. Given the size of the official GitHub schema~\cite{githubPublicSchema2024}, we restrict the graph to four entities (\emph{Users}, \emph{Commits}, \emph{Issues}, and \emph{Files}) and their relationships. These four entities capture the core workflows we evaluate (developer activity, commit history, and defect lifecycle), enabling multi-hop analysis while keeping the schema manageable for verification and traceability of responses. The knowledge graph constructor collects the following types of data:
\begin{itemize} 
    \item \textbf{\textit{Commits}}: Information about each commit to track code changes, authorship, and contributions over time. 
    \item \textbf{\textit{Issues}}: Details of issues (bugs) to track reported problems and identify their introducing and fixing commits.
    \item \textbf{\textit{Files}}: File structures and changes over time to track modifications and identify files impacted by bugs.
    \item \textbf{\textit{Users}}: Contributor information to analyze developer activities and contributions.
\end{itemize}

The knowledge graph constructor also identifies the bug-fixing and bug-introducing commits. Similar to prior studies~\cite{dacostaFrameworkEvaluatingResults2017}, our approach identifies the bug-fixing commits by searching for the bug ID in the change logs of the commits. Then it identifies the buggy changes by employing the \citet{daviesComparingTextbasedDependencebased2014} variation of the SZZ algorithm referenced as R-SZZ~\cite{dacostaFrameworkEvaluatingResults2017}. The SZZ algorithm~\cite{sliwerskiWhenChangesInduce2005} is a widely used method in software engineering for detecting bug-introducing changes. The R-SZZ variation uses textual and dependence-related changes to improve on the original SZZ algorithm~\cite{daviesComparingTextbasedDependencebased2014}.

\begin{figure}
\centering
\includegraphics[width=.9\columnwidth]{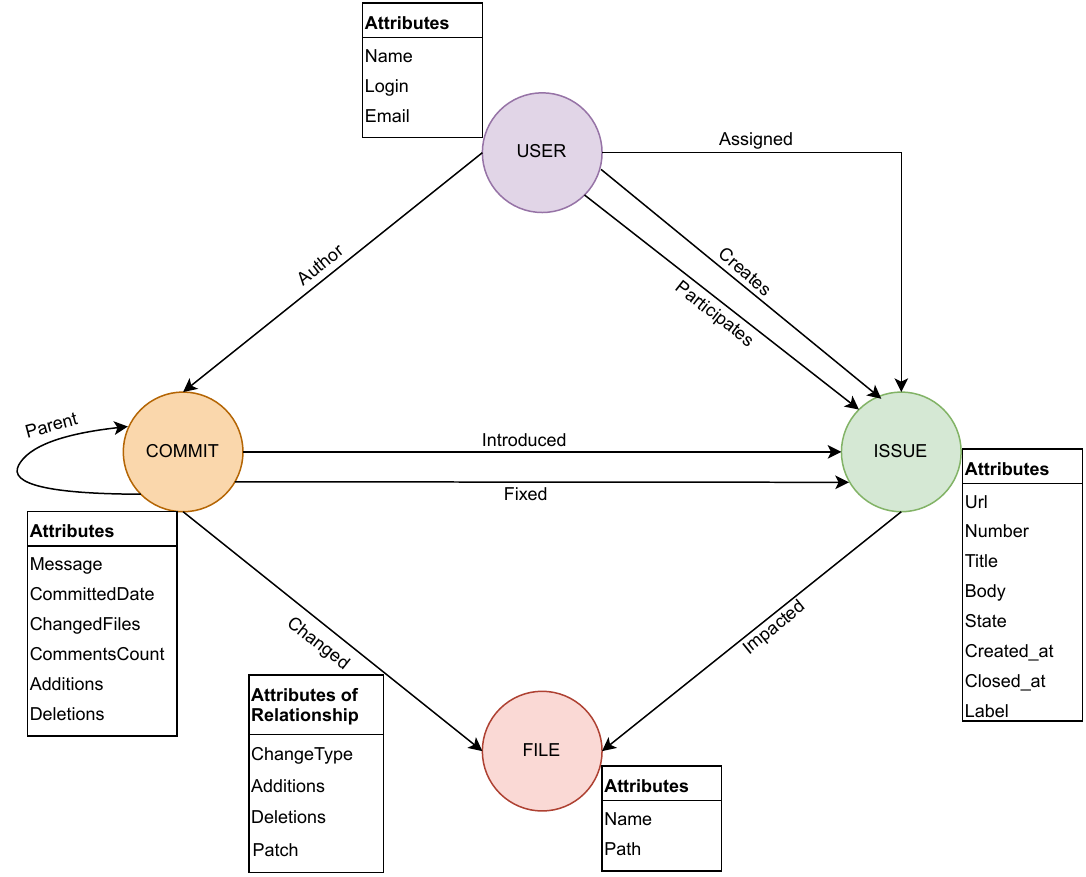}
\caption{Overview of the schema of the knowledge graph used in this study. The circles represent the entities (Nodes), the directed arrows represent the relationships (Edges), and the boxes show the attributes.}
\label{fig:knowledge_graph}
\end{figure}

After the data collection and SZZ execution, it constructs the knowledge graph. Similar to prior study~\cite{malikGraphanonymizationSoftwareAnalytics2024}, we define the schema of the knowledge graph by establishing the relationship between the entities in the GitHub repository. Figure~\ref{fig:knowledge_graph} shows an overview of the entities and relationships in the schema of our knowledge graph. A description of the relationships between the entities is presented in Table~\ref{tab:entity_prop_rel}.

For entities that continuously change during the lifespan of the repository, such as Files, we assign their evolving attributes to the relationships rather than to the nodes. For instance, if a commit changes a file, the change type (added, deleted, renamed) is assigned as an attribute to the \texttt{changed} relationship between the commit and the file, not as an attribute of the file itself.
After the construction of the knowledge graph, we store the knowledge graph in a graph database to allow for querying.

\begin{table*}
\caption{Description of relationships in our knowledge graph. The relationship between two entities is represented as (\textit{head entity}, \texttt{relationship}, \textit{tail entity})}
\label{tab:entity_prop_rel}
\centering
\begin{tabularx}{\textwidth}{l Y}
\toprule
\textbf{Relationship} & \multicolumn{1}{l}{\textbf{Description}} \\
\midrule
\rowcolor{lightgray} (\textit{User}, \texttt{Author}, \textit{Commit}) & Indicates a \textit{User} who authored a \textit{Commit}. \\
(\textit{User}, \texttt{Assigned}, \textit{Issue}) & Indicates a \textit{User} who is assigned to an \textit{Issue}.  \\
\rowcolor{lightgray} (\textit{User}, \texttt{Create}, \textit{Issue}) & Indicates a \textit{User} who created an \textit{Issue}. \\
(\textit{User}, \texttt{Participates}, \textit{Issue}) & Indicates a \textit{User} who participated in the \textit{Issue} discussion. \\
\rowcolor{lightgray} (\textit{Commit}, \texttt{Parentof}, \textit{Commit}) & Indicates that a \textit{Commit} is the parent of another \textit{Commit}. \\
(\textit{Commit}, \texttt{Introduced}, \textit{Issue}) & Indicates a \textit{Commit} that introduced or caused an \textit{Issue}. \\
\rowcolor{lightgray} (\textit{Commit}, \texttt{Fixed}, \textit{Issue}) & Indicates a \textit{Commit} that fixed an \textit{Issue}. \\
(\textit{Commit}, \texttt{Changed}, \textit{File}) & Indicates a \textit{Commit} that modified (added, deleted, renamed, modified) a \textit{File}. This relationship has properties indicating the type of change, the number of lines added to the file (additions), the number of lines deleted (deletions), and the changes (patch). \\
\rowcolor{lightgray} (\textit{Issue}, \texttt{Impacted}, \textit{File}) & Indicates an \textit{Issue} is related to or impacted the changes in a \textit{File}. \\
\bottomrule
\end{tabularx}
\end{table*}

\subsection{Query Generator}
An essential step in our approach is retrieving the relevant information to answer a user's question. The Query Generator component aims to generate graph queries that correspond to the user's questions. In this study, we operationalize the graph query using Cypher, an evolving query language for graph databases that is supported by Neo4j, Redis Graph, and Memgraph~\cite{francisCypherEvolvingQuery2018}.

\begin{figure}
\centering
\includegraphics[width=.9\columnwidth]{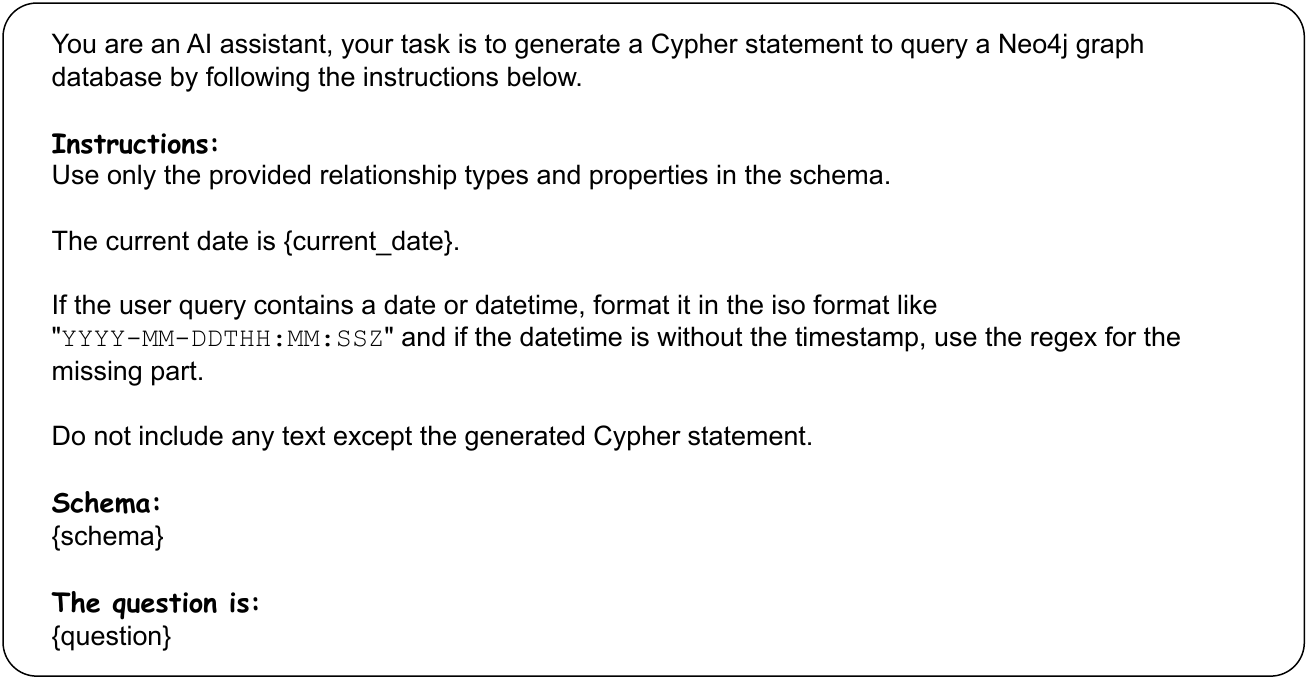}
\caption{Prompt template used by the Query Generator LLM. The prompt includes the current date and time, the schema of the knowledge graph, and the user's question.}
\label{fig:Cypher_prompt}
\end{figure}

The Query Generator uses an LLM to generate the Cypher query. The LLM uses the entities and relationships in the schema of the knowledge graph to generate the Cypher query using the prompt template shown in Figure~\ref{fig:Cypher_prompt}. The prompt follows guidelines and best practices for prompt engineering~\cite{openaiBestPracticesPrompt2024} and accepts three main parameters to generate the Cypher query: (1) the current date and time, (2) the schema of the knowledge graph, and (3) the user's natural language question. The current date and time were added to inform the LLM in answering questions requiring relative dates, such as \textit{``How many commits from last month''}. The schema of the knowledge graph informs the LLM of the types of entities and relationships in the knowledge graph.

In our running example \textit{``How many people have contributed to the code''}, the Query Generator component uses the schema of the knowledge and the question to generate the text containing the \texttt{MATCH (u:User)-[:author]->(c:Commit) RETURN COUNT(u) AS contributors} to get the number of contributors in the project.

\subsection{Query Executor}
To generate the response to the user's question, we have to retrieve and pass the relevant information from the knowledge graph to the Response Generator. We achieve this through the Query Executor, which takes the generated output of the Query Generator and executes it. Although the Query Generator is prompted to only return the Cypher query, there are instances where it returns additional texts to the Cypher query, which can result in a syntax error when executed. As a result, the Query Executor component extracts the Cypher statement from the output of the Query Generator as a means of quality control using regular expression matching.

The Query Executor then executes the extracted Cypher query, returning the results from the knowledge graph database. The result is passed to the Response Generator component to generate a natural language response for the user. In the running example, the Query Executor extracts the Cypher \texttt{MATCH (u:User)-[:author]->(c:Commit) RETURN COUNT(u) AS contributors} from the generated text. It then executes the query and returns the result [contributors: 40], assuming there are 40 contributors.

\subsection{Response Generator}
\begin{figure}
\centering
\includegraphics[width=.9\columnwidth]{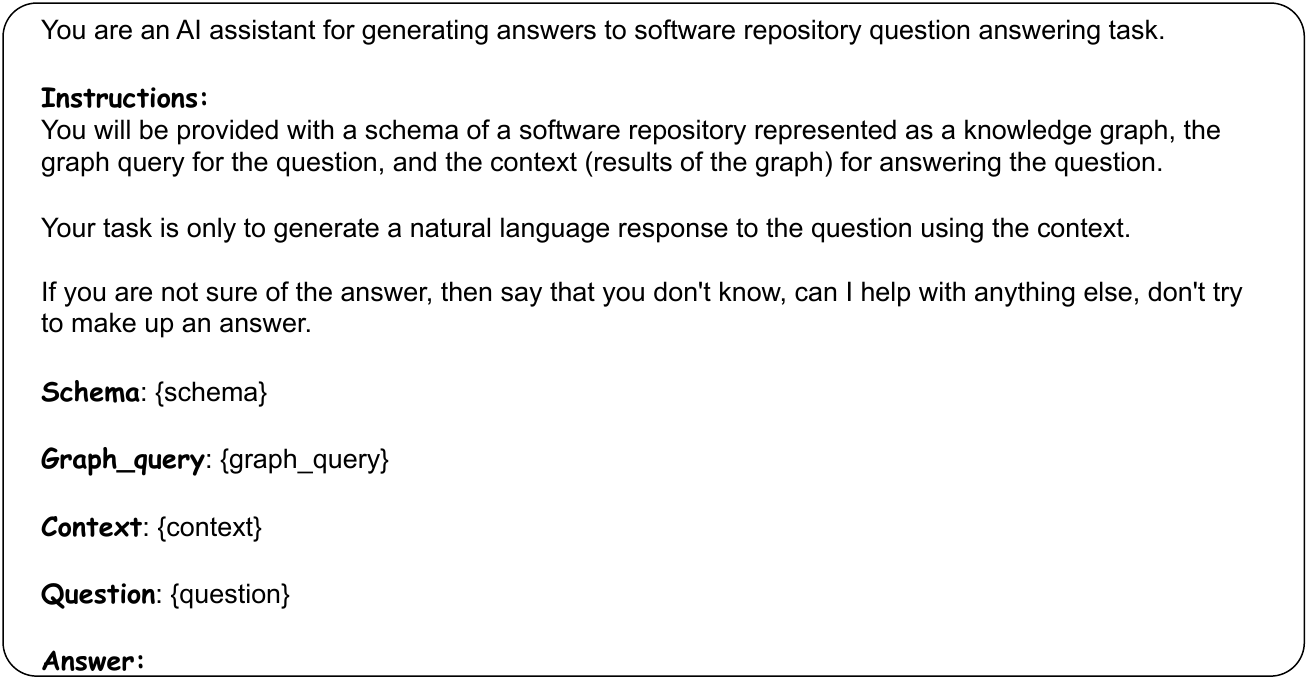}
\caption{Prompt template used by the Response Generator LLM. The prompt includes the schema of the knowledge graph, the generated Cypher query for the question, the results returned from executing the Cypher query, and the question.}
\label{fig:response_prompt}
\end{figure}

The goal of the Response Generator Component is to generate a natural language response to the user's question based on the results returned by the Query Executor. The Response Generator prompts the Response Generator LLM to generate the natural language response using the prompt template shown in Figure~\ref{fig:response_prompt}. The prompt template accepts four parameters: (1) the schema of the knowledge graph, (2) the generated Cypher query serving as additional context for interpreting the results and generating an appropriate answer to the question, (3) the context, which is the results from the Query Executor, and (4) the user's question. To prevent hallucination, we instructed the LLM to respond \textit{``I don't know''}, if it is not sure of the answer and not make up a response. In the running example, the Response Generator takes the results of the query, the question, the schema, and the Cypher query and returns the natural language response ``A total of 40 people have contributed to the code''.

\section{Evaluation setup}
\label{sec:eval_setup}
The main goal of this study is to improve the accuracy of LLM-based chatbots in answering repository-related questions. In this section, we present the evaluation setup for our approach in detail. We begin by explaining the criteria for selecting the projects for the evaluation of our approach. Finally, we discuss the questions used for the evaluation and the implementation of our approach.

\subsection{Selected Project}
\label{sec:proj_selection}
For this study, we selected software projects from GitHub based on a set of criteria for our evaluation. We selected the projects based on their popularity on GitHub. We used the number of stars of a project as a proxy for identifying the most popular projects on GitHub~\cite{borgesWhatsGitHubStar2018}. However, some of the most popular projects on GitHub are not software projects, such as a collection of awesome projects or educational projects. Therefore, we excluded projects that are not software projects, for example, the free-programming-books project~\cite{ebookfoundationEbookFoundationFreeprogrammingbooks2024}. In addition, we required that the projects have their code and issue tracking data on GitHub. This requirement ensures that all relevant development activities for the construction of the knowledge graph discussed in Section~\ref{sec:kg_constructor} are accessible through a unified platform, facilitating comprehensive data collection. For example, the Linux~\cite{torvaldsTorvaldsLinux2024} project is one of the most popular projects on GitHub, but it was excluded because the issue tracking is not on GitHub. Also, we required that for commits that are fixing or closing issues in the project, the commit log should reference the issue id, for example, \textit{``fixes issue \#123''}. This linkage is for accurately mapping issues to their fixing changes when constructing the knowledge graph and serves as a start to progressively identify the bug introducing commits~\cite{daviesComparingTextbasedDependencebased2014}.

Based on these criteria, we selected five popular open-source projects shown in Table~\ref{tab:proj_stats}. The selected projects cover various domains and programming languages and have a median number of 170,736 stars, 7,295 commits, 12,545 issues, and 392 contributors. The data were collected on August 19, 2024.

\begin{table*}
\centering
\caption{Overview of the selected projects for evaluating our approach}
\label{tab:proj_stats}
\small 
\begin{tabularx}{\textwidth}{
    >{\raggedright\arraybackslash}X 
    >{\raggedright\arraybackslash}X 
    >{\centering\arraybackslash}m{1.5cm} 
    >{\raggedright\arraybackslash}m{1.5cm} 
    >{\centering\arraybackslash}m{1.5cm} 
    >{\centering\arraybackslash}m{1cm}
    >{\centering\arraybackslash}m{1.7cm} 
    }
\toprule
\textbf{Project} & \textbf{Domain} & \textbf{Stars} & \textbf{Language} & \textbf{Commits} & \textbf{Issues} & \textbf{Contributors} \\
\midrule
\rowcolor{lightgray}React & Web framework & 225,068 & JavaScript & 19,008 & 13,090 & 416 \\
Vue & Web framework & 207,371 & TypeScript & 3,592 & 12,545 & 358 \\
\rowcolor{lightgray}Ohmyzsh & Systems utility & 170,736 & Shell & 7,295 & 12,034 & 392 \\
Bootstrap & Web framework & 168,051 & JavaScript & 22,833 & 37,634 & 361 \\
\rowcolor{lightgray}AutoGPT & AI framework & 163,726 & Python & 5,373 & 6,169 & 440 \\
\bottomrule
\end{tabularx}
\end{table*}

\subsection{Evaluation Questions}
\label{sec:eval_questions}
In evaluating the approach, we curated the questions by~\citet{abdellatifMSRBotUsingBots2020}, which they collected from 12 users interacting with software repositories to access various information for the completion of tasks assigned to them. These tasks include finding answers to questions that are commonly asked by developers and non-technical stakeholders, for instance, finding the commit that introduced a bug or the developer that fixed the most bugs~\cite{sharmaWhatDevelopersWant2017,begelAnalyzeThis1452014}. The users asked 165 questions representing 10 distinct intents, where each intent refers to the mapping between the user's question and a predefined action to be taken to complete the task~\cite{adamopoulouOverviewChatbotTechnology2020}. 

Out of the 165 questions in the dataset, we exclude 15 questions that are not relevant to the evaluation. For example, \textit{``What’s your name?''}. We use the remaining questions as templates for forming the evaluation questions for each project. This involves inserting project-specific parameters into the question template to form the question. For instance, in the original dataset, for the question \textit{``Who fixed the most bugs in the file HibernateEntityManager?''}, we replace the file \texttt{HibernateEntityManager} with a file specific to the project we are evaluating. Executing all 150 questions for each project would be expensive; therefore, we create a subset of the dataset by selecting two questions per intent, resulting in an evaluation set of 20 question templates. The two questions were selected for each intent based on the clarity of their phrasing. This approach reduces the cost of executions while ensuring variety by covering each of the 10 intents with two variations of questions.

Also, for a more fine-grained evaluation of our approach, we classify the selected questions into three difficulty levels. We define difficulty as the number of relationships in the knowledge graph required to answer the question. The level one questions include questions that only require a single entity and not a relationship to answer. For example, \textit{``What is the latest commit''} only requires the \textit{Commit} entity. Level two questions require a single relationship to answer. For example, \textit{``Which commit fixed the bug X''} requires the \textit{Commit} entity and the \textit{Issue} entity linked by the \texttt{fixed} relationship. Lastly, the level three questions require two or more relationships to answer. For example, \textit{``Determine the percentage of fixing commits that introduced bugs in June 2018''} requires the \textit{Commit} entity, the \textit{Issue} entity, the \texttt{fixed} relationship and \texttt{introduced} relationship. The 20 questions used for the evaluation with their corresponding intent and difficulty level can be found in Appendix~\ref{appendix:a}.

To establish the ground truth for our evaluation, the first author manually wrote Cypher queries corresponding to all 20 questions for each of the selected repositories. To ensure the correctness of these queries and eliminate potential bias, the authors collaboratively reviewed and discussed the logic employed in each query, adding an additional layer of scrutiny. The Cypher queries were then executed against the knowledge graphs, and the resulting outputs were used as the ground truth for comparison in our study.






\begin{table*}[t]
\centering
\caption{Definition of difficulty levels along with example questions and corresponding Cypher queries}
\label{tab:question_group}
\footnotesize
\setlength{\tabcolsep}{6pt}
\renewcommand{\arraystretch}{1.1}
\begin{tabularx}{\textwidth}{p{0.8cm}p{2.5cm}p{0.5cm}p{3.0cm}X}
\toprule
\textbf{Level} & \textbf{Definition} & \textbf{\#} & \textbf{Example} & \textbf{Cypher Query} \\
\midrule

\rowcolor{lightgray}
1 & Questions requiring only a single entity, no relationship needed & 4 &
What is the latest commit? &
\codeblock{%
MATCH (c:Commit)\\
RETURN c\\
ORDER BY c.committedDate DESC\\
LIMIT 1
}
\\ 
\RowGap

2 & Questions requiring one relationship & 12 &
Determine the developers that had the most unfixed bugs? &
\codeblock{%
MATCH (u:User)-[:assigned]->(i:Issue)\\
WHERE i.state = "open"\\
RETURN u, COUNT(i) AS openBugs\\
ORDER BY openBugs DESC
}
\\ 
\RowGap

\rowcolor{lightgray}
3 & Questions requiring two or more relationships & 4 &
Determine the developers that fixed the most bugs in ReactDOMInput.js? &
\codeblock{%
MATCH (u:User)-[:author]->(c:Commit)\\
        -[:fixed]->(i:Issue)\\
      -[:impacted]->(f:File \{name: "ReactDOMInput.js"\})\\
RETURN u, COUNT(i) AS fixedBugs\\
ORDER BY fixedBugs DESC
}
\\

\bottomrule
\end{tabularx}
\end{table*}

\subsection{Implementation}
We implement the approach discussed in Section~\ref{sec:approach} using Python and the Langchain framework.
The Knowledge Graph Constructor begins the process by collecting data from the software repository using the GitHub GraphQL API~\cite{githubGitHubGraphQLAPI2024}. We opted for the GraphQL API over the REST API~\cite{githubGitHubRESTAPI2024} because GraphQL allows us to specify precisely the data we need in a single request, reducing the noise and the size of the document returned compared to the REST API and improving processing efficiency~\cite{britoMigratingGraphQLPractical2019}. The collected data included information on users, commits, issues, and files (see Section~\ref{sec:kg_constructor}). After collecting the data, we implemented the relationship between the entities following the schema defined in Figure~\ref{fig:knowledge_graph}. To store and manage the knowledge graph, we utilize the Neo4j database, which is known for its robustness and maturity in handling graph data structures and widely adopted in prior studies~\cite{malikGraphanonymizationSoftwareAnalytics2024}. Its compatibility with the Cypher query language enables efficient querying and manipulation of the graph data~\cite{francisCypherEvolvingQuery2018}.

In the Query Generator and Response Generator components, we use OpenAI's GPT-4o model through OpenAI's API to translate the user's natural language questions into Cypher queries and also generate user-friendly responses. The selection of the GPT-4o model is based on its performance in our exploratory question (discussed in Section~\ref{rq0}). For the implementation of our approach, we use the default setting of the model except in the Query Generator component, where we set the temperature of the model to 0. We use temperature 0 to reduce the randomness in the generated Cypher queries and main consistency~\cite{liCanLLMAlready2023}.

After the Cypher query is generated, it is executed using the Neo4j Python library, which provides a straightforward interface for communicating with the Neo4j database. This library enables the approach to run the query and retrieve the relevant results from the knowledge graph efficiently for response generation.

\subsection{Baseline Set Up}
\label{sec:baseline_setup}
We benchmark our approach against two baseline approaches, the MSRBot framework~\cite{abdellatifMSRBotUsingBots2020} and the GPT-4o model with web searching capability.
The MSRBot framework is an intent-based chatbot framework that maps user questions to predefined intents and also extracts the relevant entities from the question to retrieve the relevant repository data. In the implementation of MSRBot, Google Dialogflow was used for natural language understanding. However, the authors explained that other NLU platforms can be used in the framework and further evaluated different NLU platforms~\cite{abdellatifComparisonNaturalLanguage2022}. In the study, \citet{abdellatifComparisonNaturalLanguage2022} found that Rasa NLU produces stable confidence scores (median > 0.91) and generally outperforms Dialogflow in confidence reliability. As a result, in this study we implement the MSRBot framework with the Rasa NLU.

The second baseline approach we consider is an open-domain LLM-based chatbot that can handle repository-related questions by augmenting a large language model with retrieval. A common method for this is RAG. The RAG approach integrates an LLM with a knowledge base: user queries trigger a search in an indexed repository dataset, and the retrieved documents are then used by the LLM to generate an answer. In principle, RAG allows the bot to provide up-to-date, factual answers even about dynamic repository data. We initially considered a RAG-based baseline, similar to the one explored by~\citet{abeduLLMBasedChatbotsMining2024} for mining software repositories. However, we do not include the RAG baseline in our evaluation. The reason is that the RAG chatbot showed poor accuracy, failing to answer most of the questions in the MSRBot dataset. The authors explained the low accuracy as a result of retrieving irrelevant data or failing to generate correct answers. This outcome highlights that a naive RAG implementation is ineffective for this evaluation, as reported by~\citet{abeduLLMBasedChatbotsMining2024}. Given this finding, a RAG baseline would not be a competitive baseline. Instead, we replace the RAG approach with an LLM with web-search capability baseline: GPT-4o with web search capabilities (the GPT-4o-search-preview model). This baseline uses OpenAI’s GPT-4o, which is a state-of-the-art LLM, augmented with the ability to perform live web searches. The GPT-4 model can thus retrieve current information from online sources (including public GitHub repositories) and incorporate that knowledge into its responses. This serves a similar purpose to the RAG approach, providing the language model with access to relevant up-to-date information.

\subsection{Survey Design}
\label{sec:survey_design}
To evaluate the usefulness of our approach in assisting users with repository-related questions, we designed a task-based user study. This study follows a methodology similar to that of~\citet{abdellatifMSRBotUsingBots2020}. Participants are asked to perform a set of software repository-related tasks: first, using conventional tools (baseline), and then using our chatbot-based approach. By comparing the performance of the tasks in these two settings, we assess the usefulness of our approach for task accuracy and efficiency.

We adopted the 10 tasks performed by participants in~\citet{abdellatifMSRBotUsingBots2020}. These tasks cover a range of questions developers ask about the repositories. To avoid overburdening participants, we divided the ten tasks into two sets of five. We randomly selected two projects (Vue and Ohmyzsh) from our study projects discussed in Section~\ref{sec:proj_selection} as the subjects for these tasks. Participants were then randomly split into two groups (Group A and Group B). Group A was assigned the first set of five tasks on the Vue project, and Group B was assigned the second set of five tasks on the Ohmyzsh project. The groupings helped ensure that no participant had to perform all ten tasks and that each participant only dealt with one project’s context, reducing fatigue and learning effects.

We developed and deployed a web-based chatbot application\footnote{\url{https://repochattool.streamlit.app/}}
 implementing our approach, which participants used to interact with the projects. The study was administered via an online survey (Qualtrics), structured into four sections. The first section collects demographic information about the participants. This offers us insight into the professional background, the number of years of experience the participants have using Git tools, and how often they use the Git tools.

In the second section, participants were presented with five tasks and instructed to complete each task using any tools or methods of their choice (except our chatbot). They could use Git command-line queries, repository browsing, writing scripts, web searches, or AI assistants, as desired. In order for participants not to spend all their effort on a single task and keep within the allotted time of 60 minutes for the survey, we asked the participants to spend not more than 6 minutes per task (i.e., approximately 30 minutes total for the five tasks in this section). After completing each task, the participants provided their answer (the outcome of the task) and a brief description of how they completed the task. The description allowed us to understand the effort and techniques adopted by the participants to complete the tasks. We did not impose strict tool restrictions in order to emulate a realistic scenario where developers can leverage all resources at their disposal. This baseline phase established a point of comparison for task difficulty and time taken without the assistance of our tool.

In the third section, participants performed the same set of five tasks, but this time using our chatbot. We asked the participants to phrase each task as a question in their own words to the chatbot. For each task, the survey asked them to copy the exact question they posed to the chatbot and the answer the chatbot returned. After receiving the chatbot’s answer, participants were shown the expected correct answer (ground truth) for that task and asked to evaluate the chatbot’s response. They indicated whether the bot’s answer was Correct, Partially Correct, or Incorrect, and could optionally provide comments explaining their choice. We captured the time taken to complete each task with the chatbot as well. This section allowed us to assess the chatbot’s effectiveness in answering the questions directly compared to the baseline.

In the final section, participants answered a brief questionnaire about their overall experience with the chatbot. They rated, on a 5-point Likert scale, the chatbot’s accuracy, usefulness, time-saving capability, and their overall satisfaction with the chatbot (1 = Strongly Disagree to 5 = Strongly Agree for each statement). They were also given an open-ended text box to provide any additional feedback, comments, or suggestions about the chatbot. This feedback provides insights into the user-perceived benefits or issues with our approach.

The survey was configured to automatically record the time spent on each task (using Qualtrics’ timing features), giving us precise measurements for task completion times in both the baseline and chatbot settings.
We piloted the survey with two participants. We used the feedback from these participants to improve and refine the survey for clarity before administering it to the other participants. For the distribution of the survey, we reached out to our network of contacts from both academia and industry to participate. We also adopted a snowball technique by asking the participants to share with other professionals with Git experience in their network.

\section{Results}
\label{sec:results}
In this section, we first present the results of the exploratory analysis, and then present the results of the four main research questions. For each research question, we present the motivation, the approach, and the results.

\subsection{RQ0: \rqz}
\label{rq0}

\noindent\textbf{Motivation.} Before answering the research questions in this study, we first evaluate the capability of different LLMs to accurately generate Cypher queries from natural language text. This RQ aims to empirically assess the context of this study (i.e., the ability of LLMs to generate accurate Cypher queries from natural language text for retrieving data from a knowledge graph). Secondly, this RQ aims to empirically identify the most efficient LLM in generating Cypher queries, which will serve as the LLM model in our implementation to answer the remaining RQs.

\smallskip
\noindent\textbf{Approach.} To evaluate the capability of LLMs to generate valid Cypher queries from natural language text, we selected three state-of-the-art models considering both open-source and closed-source options: GPT-4o~\cite{openaiHelloGPT4o2024}, Llama3-8B~\cite{metaMetallamaMetaLlama38BHugging2024}, and Claude3.5~\cite{anthropicIntroducingClaude352024}. These models have been widely used in software engineering literature~\cite{zhouBridgingDesignDevelopment2024,majdoubDebuggingOpenSourceLarge2024}

Similar to the strategy by~\citet{liCanLLMAlready2023}, we evaluate the models under zero-shot settings to assess their generalization ability to generate Cypher queries using their pre-existing knowledge without prior exposure or clues. We evaluate the models on the 20 questions described in Section~\ref{sec:eval_questions} using the prompt template shown in Figure~\ref{fig:Cypher_prompt} to generate the Cypher query. Due to the stochastic nature of LLMs, we run the experiments five times, each time on a different repository~\cite{edgeLocalGlobalGraph2024}.

As our evaluation metric, we use the Execution Accuracy (EX)~\cite{liCanLLMAlready2023} because it measures the correctness of the generated Cypher queries in terms of their execution results, which is essential for applications that require accurate retrieval of data. EX is defined in Equation~\ref{eqn:1} as the proportion of the evaluation set in which the executed results of the generated queries are similar to the ground truth, relative to the examples in the evaluation set and formalized as:

\begin{equation}
\label{eqn:1}
EX = \frac{\sum_{n=1}^{N} \mathbbm{1}(V_n, \hat{V}_n)}{N}
\end{equation}
and \( \mathbbm{1}(\cdot) \) is a function represented as:

\[
\mathbbm{1}(V_n, \hat{V}_n) =
\begin{cases} 
1, & \text{if } V_n = \hat{V}_n \\
0, & \text{if } V_n \neq \hat{V}_n
\end{cases}
\]
where $N$ is the total number of executions, $V_n$ is the result set from executing the ground-truth Cypher queries, and $\hat{V}_n$ is the result set from executing the generated Cypher queries.

\smallskip
\noindent\textbf{Results.} Table~\ref{tab:ex_models} compares the execution accuracy of the selected models in generating Cypher queries. We find that GPT-4o is the most efficient model in translating natural language questions into accurate Cypher queries within the given zero-shot setting, achieving an EX score of 0.65. The superior performance of GPT-4o can be attributed to the advanced language understanding capabilities of the GPT-4 family of LLMs in capturing the semantic details required for precise Cypher query generation as demonstrated in prior studies~\cite{liCanLLMAlready2023}. The performance difference between GPT-4o, Claude3.5, and Llama3 highlights the variability in capability among different LLMs when applied to the task of generating Cypher queries from natural language text. This finding informs our decision to utilize GPT-4o for the subsequent research questions, as it offers the most reliable performance for synergizing LLMs with knowledge graph data.

\begin{table*}
\centering
\caption{Comparison of three state-of-the-art LLMs in terms of execution accuracy (EX) in generating valid Cypher queries from natural language text}
\label{tab:ex_models}
\begin{tabular}{lcccc}
\toprule
\textbf{Model} & \textbf{Questions} & \textbf{Executions} & \textbf{Correct} & \textbf{EX} \\
\midrule
\rowcolor{lightgray} GPT-4o     & 20 & 100 & 65 & 0.65 \\
Claude3-5  & 20 & 100 & 61 & 0.61 \\
\rowcolor{lightgray} Llama3     & 20 & 100 & 20 & 0.20 \\
\bottomrule
\end{tabular}
\end{table*}

\bigskip
\begin{tcolorbox}
    \paragraph{\emph{\textbf{RQ0 Summary:}}} LLMs have the capability to capture semantic details for Cypher query generation from natural language text. Specifically, GPT-4o demonstrated to be the most efficient in the task of generating Cypher queries from natural language text.
\end{tcolorbox}
\medskip

\subsection{RQ1: \rqi}
\label{rq1}

\noindent\textbf{Motivation.} The synergy between knowledge graphs and large language models (LLMs) has the potential to enhance the ability of LLMs to provide accurate and contextually relevant answers~\cite{panUnifyingLargeLanguage2024}. Knowledge graphs encapsulate structured information about entities and their relationships, which can be crucial for understanding complex queries and providing precise answers. In this research question, we investigate the effectiveness of generating an accurate response to a user question by adding a layer of semantic understanding. This enables the LLM to generate a Cypher query and retrieve the relevant information to answer the question.

\smallskip
\noindent\textbf{Approach.}
To evaluate the performance of our approach in answering software repository-related questions, we conducted an end-to-end evaluation of the approach from the moment the Query Generator receives the natural language query to when the Response Generator outputs the final response (see step 2 in Figure~\ref{fig:approach_overview}). The end-to-end evaluation measures the practical performance of our approach in generating accurate answers~\cite{edgeLocalGlobalGraph2024}.

For this purpose, we evaluated the 20 questions described in Section~\ref{sec:eval_questions} for each of the selected projects. 
For each question, we also executed the process five times to account for the stochastic nature of LLMs in the generation process.
We compared the final responses generated by our approach to the oracle answers (predetermined correct answers based on the data in the knowledge graph). A question was considered correctly answered if our approach provided the correct response in at least 3 out of 5 executions. If it failed in three or more executions, the question was marked as incorrect.

\smallskip
\noindent\textbf{Results.}
Table~\ref{tab:repo_comp_rq1} compares the accuracy of our approach across the selected projects. We observed that the accuracy of our approach varies between 60\% and 75\% across the projects. The highest accuracy was achieved for the AutoGPT repository at 75\%, while both Ohmyzsh and Vue had the lowest accuracy at 60\%. The average accuracy across all repositories was 65\%, indicating that our approach has a moderate overall effectiveness in answering questions related to the software repositories.

Table~\ref{tab:level_comp_rq1} also compares the accuracy of our approach based on the difficulty level of the questions. The results show a correlation between the difficulty level and the accuracy of our approach. Our approach achieved an 80\% accuracy on level 1 questions, and the accuracy decreased to 65\% for level 2 questions. The accuracy further dropped to 50\% for level 3 questions. This trend suggests that while our approach is effective at handling straightforward queries, its effectiveness decreases as the questions get more complex. We present all the questions answered correctly or incorrectly in this RQ in Appendix~\ref{appendix:e}

\begin{table*}
\centering
\caption{Comparison of the accuracy of our approach across the selected projects}
\label{tab:repo_comp_rq1}
\begin{tabular}{lcccc}
\toprule
\textbf{Project} & \textbf{Questions} & \textbf{Answered} & \textbf{Correct} & \textbf{Accuracy} \\
\midrule
\rowcolor{lightgray} AutoGPT    & 20 & 20 & 15 & 0.75  \\
Bootstrap  & 20 & 17 & 13 & 0.65 \\
\rowcolor{lightgray} Ohmyzsh    & 20 & 14 & 12 & 0.60  \\
React      & 20 & 18 & 13 & 0.65 \\
\rowcolor{lightgray} Vue        & 20 & 17 & 12 & 0.60  \\
\midrule
\textbf{Overall}  & \textbf{100} & \textbf{86} & \textbf{65} & \textbf{0.65} \\
\bottomrule
\end{tabular}
\end{table*}

\begin{table*}
\centering
\caption{Comparison of the accuracy of our approach based on the difficulty level of the questions}
\label{tab:level_comp_rq1}
\begin
{tabular}{lcccc}
\toprule
\textbf{Level} & \textbf{Questions} & \textbf{Answered} & \textbf{Correct} & \textbf{Accuracy} \\
\midrule
\rowcolor{lightgray} 1     & 20  & 16  & 16  & 0.80  \\
2     & 60  & 51  & 39   & 0.65  \\
\rowcolor{lightgray} 3     & 20  & 19  & 10  & 0.50 \\
\midrule
\textbf{Total} & \textbf{100} & \textbf{86} & \textbf{65} & \textbf{0.65} \\
\bottomrule
\end{tabular}
\end{table*}

\bigskip
\begin{tcolorbox}
    \paragraph{\emph{\textbf{RQ1 Summary:}}} Synergizing LLMs with knowledge graphs correctly answered repository-related questions 65\% of the time. The approach performs well when answering simple questions but struggles with complex questions that require two or more relationships from the knowledge graph.
\end{tcolorbox}
\medskip

\subsection{RQ2: \rqii}
\label{rq2}

\noindent\textbf{Motivation.} While our approach achieved better performance compared to previous LLM-based approaches~\cite{abeduLLMBasedChatbotsMining2024}, the task is still challenging for our approach, with an accuracy of 65\%. To be able to improve our approach to achieve a higher accuracy, we need to understand the reasons for which our approach fails. Therefore, in this research question, our goal is to identify the limitations of our approach by manually analyzing the incorrectly generated responses of our approach.

\smallskip
\noindent\textbf{Approach.} To understand the limitations of our approach, we selected all the incorrectly answered questions from the 500 executions in RQ1, that is, 20 questions each executed five times for five repositories. We identified and manually analyzed a total of 164 executions that returned incorrect answers. To identify the limitations, we adopted an open-card sorting approach as used in prior studies~\cite{latendresseExploratoryStudyMachine2024}. The executions were sorted based on the final response of our approach, the generated Cypher query, and the results from executing the Cypher query. The main author read through all 164 executions to identify recurring themes and patterns that may have led to incorrect responses to come up with labels. To ensure the labels are less biased and go through a level of scrutiny, the authors discussed each question and the preliminary labels. This step ensured that the labels had clarity and were relevant. Based on this step, some of the labels were merged, split, or modified to provide more clarity.

\smallskip
\noindent\textbf{Results.}
Table~\ref{tab:limitations} summarizes the definitions, frequencies, and percentages of the six main limitations we identified from manually analyzing the instances where our approach generated incorrect answers: Incorrect relationship modeling, Faulty arithmetic logic, Misapplied attribute filtering, Invalid assumptions, Misapplied date formatting, and Hallucination. There were cases where multiple limitations were identified within a single instance. Thus the frequency reported reflects the total number of limitations rather than the 164 instances analyzed.

The most prevalent limitation was the \textbf{incorrect relationship modeling}, occurring in 123 out of 164 incorrect responses (75.0\%). This limitation occurs when the logic used by the LLM to generate the query deviates from the intention of the question. This limitation is due to the interpretation the LLM places on the question; if the interpretation is incorrect, the LLM will proceed to generate a logically incorrect query. For instance, as shown in Table~\ref{tab:reasoning_eg}, consider the question: \textit{``Determine the percentage of the fixing commits that introduced bugs on July 2023"}. The correct Cypher query first matches all commits that fixed an issue in July 2023 to find the total fixing commits. It then matches the commits that both fixed an issue and introduced another issue in July 2023 and calculates the percentage accordingly. In contrast, the LLM generated a syntactically correct but logically flawed Cypher query. It incorrectly matched commits that fixed an issue as the fixing commits and separately matched commits that introduced an issue as the introducing commits without linking the two actions within the same commits. This misinterpretation of the relationships leads to an incorrect result.

\textbf{Faulty arithmetic logic} was observed in 19 instances, signifying 11.6\% of the total instances analyzed. This limitation deals with instances where the LLM is required to perform an arithmetic operation to return a final response to the user but fails to perform the correct arithmetic operation. For instance, in the example in Table~\ref{tab:reasoning_eg}, the fixingCommits should be the denominator when calculating a percentage of fixing commits. However, in the generated query by the LLM, the introducing commit was used as the denominator in the context, which is incorrect.

Also, in 16 instances (9.8\%), we found the LLM \textbf{misapplying the attributes when filtering} the matched data. In these instances, the LLM used the wrong attribute to filter the data based on the constraints specified in the question. The example in Table~\ref{tab:filtering} shows the LLM correctly matching commits that introduced bugs. However, when filtering to the specified date in the question, the LLM misapplied the filter to the issue creation date instead of the commit date. This error by the LLM led to generating an incorrect final response.

\textbf{Making Incorrect Assumptions} was also observed in 15 instances (9.1\%) of the incorrectly answered executions. In these cases, the LLM generated a syntactically correct query but made incorrect assumptions that led to an incorrect answer. For example, as shown in Table~\ref{tab:assumption_eg}, when asked \textit{``Return a commit message on July 31?"}, the LLM assumed a specific year \texttt{``2024-07-31''} in the query, whereas the correct approach would be to use a wildcard for the year and match any date that includes \texttt{``-07-31''} to retrieve commits on July 31 of any year. This incorrect assumption limits the query to a specific date, potentially excluding relevant results. If there's no record for the assumed date, then LLM will respond incorrectly.

We also identified \textbf{misapplied date formatting} in 10 instances (6.1\%). This refers to cases when the format of the data in the query does not reflect the format accepted by the graph database or the formatting incorrectly filters out data that are not to be filtered out. For instance, in the example in Table~\ref{tab:date}, the query generated by the LLM unintentionally filters out data due to the formatting of the date. These filtered data will be returned if the LLM uses a wildcard.

\textbf{Hallucination} was another limitation, occurring in 10 instances (6.1\%). This happens when the LLM makes use of nodes and relationships that are not present in the knowledge graph schema. As illustrated in Table~\ref{tab:hallucination_eg}, for the question \textit{``Determine the developers that fixed the most bugs in \texttt{challenge.py}?''}, the LLM introduced a non-existent relationship \texttt{[fixed]} directly between the \texttt{User} and \texttt{Issue} nodes. In the correct schema, the \texttt{[fixed]} relationship exists between the \texttt{Commit} and \texttt{Issue} nodes, not between \texttt{User} and \texttt{Issue}. This hallucination leads to an invalid query and incorrect results.

The limitation on incorrectly modeling relationships, faulty arithmetic logic, misapplied attribute filtering, and date formatting highlights the limitation in the reasoning ability of the LLM. This hinders the LLM in accurately interpreting and utilizing the right nodes and relationships within the knowledge graph. This aligns with prior studies that highlight the challenges LLMs face in reasoning tasks~\cite{chenImprovingLargeLanguage2024,wangLargeLanguageModels2023,wangMakingLargeLanguage2023}. The incorrect reasoning often leads to misconstructed queries that do not align with the intended question, resulting in inaccurate or irrelevant answers. Addressing these limitations is crucial for improving our approach. By refining the LLM's understanding of the knowledge graph schema and enhancing its reasoning capabilities, we can reduce the incidence of incorrect responses.

\begin{table*}
\centering
\caption{Summary of the limitations identified with their frequency and percentage (the sum of the frequencies is more than 164 because we identified multiple limitations in some instances)}
\label{tab:limitations}
\footnotesize
\begin{tabularx}{\textwidth}{>{\raggedright\arraybackslash}X >{\raggedright\arraybackslash}X c c}
\toprule
\textbf{Limitation} & \textbf{Definition} & \textbf{Frequency} & \textbf{Percentage} \\
\midrule
\rowcolor{lightgray} Incorrect relationship modelling & The LLM deviates from the intended logic of the question, leading to a query that incorrectly models relationships within the knowledge graph & 123 & 75.0\% \\
Faulty arithmetic logic & The LLM fails to perform correct arithmetic operations required to generate the final response & 19 & 11.6\% \\
\rowcolor{lightgray} Misapplied attribute filtering & The LLM uses incorrect attributes to filter data, not adhering to the constraints specified in the question & 16 & 9.8\% \\
Invalid assumptions & The LLM makes incorrect assumptions, such as assuming specific values not provided, leading to incorrect or limited results & 15 & 9.1\% \\
\rowcolor{lightgray} Misapplied date formatting & The LLM uses incorrect date formats, not matching the database format and improperly filtering out data & 10 & 6.1\% \\
Hallucination & The LLM introduces nodes or relationships not present in the knowledge graph schema, resulting in invalid queries & 10 & 6.1\% \\
\bottomrule
\end{tabularx}
\end{table*}

\begin{table*}
\centering
\caption{Example of incorrect relationship modeling and faulty arithmetic logic in the generated query}
\label{tab:reasoning_eg}
\footnotesize
\begin{tabular}{p{6.5cm}p{6.5cm}}
\toprule
\textbf{Correct Query} & \textbf{Generated Incorrect Query} \\
\midrule
\textbf{Q: Determine the percentage of the fixing commits that introduced bugs on July 2023?} & \textbf{Q: Determine the percentage of the fixing commits that introduced bugs on July 2023?} \\
\begin{lstlisting}[style=greenHighlight]
@MATCH (c:Commit)-[:fixed]->(i:Issue)@
WHERE c.committedDate =~ "2023-07.*"
WITH COUNT(c) AS totalFixingCommits
@MATCH (i2:Issue)<-[:introduced]-(c:Commit)-[:fixed]->(i:Issue)@
WHERE c.committedDate =~ "2023-07.*"
WITH totalFixingCommits, COUNT(c) AS fixingAndIntroducingBugs
@RETURN (toFloat(fixingAndIntroducingBugs) / totalFixingCommits) * 100 AS percentage@
\end{lstlisting} & 
\begin{lstlisting}[style=redHighlight]
@MATCH (c1:Commit)-[:fixed]->(:Issue)<-[:introduced]-(c2:Commit)@
WHERE c1.committedDate =~ '2023-07.*' AND c2.committedDate =~ '2023-07.*'
WITH count(c1) AS fixingCommits, count(c2) AS introducingCommits
@RETURN (toFloat(fixingCommits) / introducingCommits) * 100 AS percentage@
\end{lstlisting} \\
\bottomrule
\end{tabular}
\end{table*}

\begin{table*}
\centering
\caption{Example of misapplied attribute filtering in the generated query}
\label{tab:filtering}
\footnotesize
\begin{tabular}{p{6.5cm}p{6.5cm}}
\toprule
\textbf{Correct Query} & \textbf{Generated Incorrect Query} \\
\midrule
\textbf{Q: What commits were buggy on June 08, 2023?} & \textbf{Q: What commits were buggy on June 08, 2023?} \\
\begin{lstlisting}[style=greenHighlight]
MATCH (c:Commit)-[:introduced]->(i:Issue)
@WHERE c.committedDate =~ '2023-06-08T.*'@
RETURN c
\end{lstlisting} & 
\begin{lstlisting}[style=redHighlight]
MATCH (c:Commit)-[:introduced]->(i:Issue)
@WHERE i.created_at =~ '2023-06-08T.*'@
RETURN c
\end{lstlisting} \\
\bottomrule
\end{tabular}
\end{table*}

\begin{table*}
\centering
\caption{Example of invalid assumptions in the generated query}
\label{tab:assumption_eg}
\footnotesize
\begin{tabular}{p{6.5cm}p{6.5cm}}
\toprule
\textbf{Correct Query} & \textbf{Generated Incorrect Query} \\
\midrule
\textbf{Q: Return a commit message on July 31?} & \textbf{Q: Return a commit message on July 31?} \\
\begin{lstlisting}[style=greenHighlight]
MATCH (c:Commit)
@WHERE c.committedDate =~ ".*-07-31T.*"@
RETURN c.message
\end{lstlisting} & 
\begin{lstlisting}[style=redHighlight]
MATCH (c:Commit)
@WHERE c.committedDate =~ '2024-07-31T.*'@
RETURN c.message
\end{lstlisting} \\
\bottomrule
\end{tabular}
\end{table*}

\begin{table*}
\centering
\caption{Example of misapplied date formatting in the generated query}
\label{tab:date}
\footnotesize
\begin{tabular}{p{6.5cm}p{6.5cm}}
\toprule
\textbf{Correct Query} & \textbf{Generated Incorrect Query} \\
\midrule
\textbf{Q: How many fixing commits caused bugs on July 2023?} & \textbf{Q: How many fixing commits caused bugs on July 2023?} \\
\begin{lstlisting}[style=greenHighlight]
MATCH (c:Commit)-[:fixed]->(:Issue), 
(c)-[:introduced]->(:Issue)
@WHERE c.committedDate =~ '2023-07-.*'@
RETURN count(c) AS fixingIntroducingCommits
\end{lstlisting} & 
\begin{lstlisting}[style=redHighlight]
MATCH (c:Commit)-[:fixed]->(:Issue)
@WHERE c.committedDate =~ '2023-07-..T..:..:..Z'@
RETURN count(c) AS fixingCommits
\end{lstlisting} \\
\bottomrule
\end{tabular}
\end{table*}

\begin{table*}
\centering
\caption{Example of hallucination in the generated query}
\label{tab:hallucination_eg}
\footnotesize
\begin{tabular}{p{6.5cm}p{6.5cm}}
\toprule
\textbf{Correct Query} & \textbf{Generated Incorrect Query} \\
\midrule
\textbf{Q: Determine the developers that fixed the most bugs in challenge.py?} & \textbf{Q: Determine the developers that fixed the most bugs in challenge.py?} \\
\begin{lstlisting}[style=greenHighlight]
@MATCH (u:User)-[:author]->(c:Commit)-[:fixed]->(i:Issue)-[:impacted]->(f:File {name: "challenge.py"})@
RETURN u, COUNT(i) AS fixedBugs
ORDER BY fixedBugs DESC
\end{lstlisting} & 
\begin{lstlisting}[style=redHighlight]
@MATCH (u:User)-[:fixed]->(i:Issue)-[:impacted]->(f:File {name: "challenge.py"})@
RETURN u.name AS developer, COUNT(i) AS bugs_fixed
ORDER BY bugs_fixed DESC
LIMIT 1
\end{lstlisting} \\
\bottomrule
\end{tabular}
\end{table*}

\bigskip
\begin{tcolorbox}
    \paragraph{\emph{\textbf{RQ2 Summary:}}} 
    The reasoning of LLM during query generation is the most prevalent limitation affecting LLMs when enhanced with knowledge graphs. It hinders the ability of the LLM to accurately interpret and utilize the right nodes and relationships within the knowledge graph, leading to incorrect relationship modeling, faulty arithmetic logic, misapplied attribute filtering, and misapplied date formatting. Other limitations include the LLM making wrong assumptions and hallucination.
\end{tcolorbox}
\medskip

\subsection{RQ3: \rqiii}
\label{rq3}

\noindent\textbf{Motivation.} 
In RQ2, the failure analysis showed that most of the limitations leading to an incorrect response are due to the \emph{faulty reasoning of the LLM during query generation}. To address this limitation, we introduce \emph{chain-of-thought (CoT)} prompting in the LLM underlying the \textbf{Query Generator}. Our goal for this research question is to improve the reasoning ability of the LLM in the \textbf{Query Generator} to minimize faulty reasoning. We do this by introducing chain-of-thought (CoT) prompting into our approach, feeding the LLM step-by-step reasoning examples to guide it in generating a reasoning path to reach an answer. Prior studies have shown that using chain-of-thought (CoT) to generate a series of intermediate steps before the final answer can improve the reasoning ability of LLMs~\cite{kojimaLargeLanguageModels2022, weiChainofThoughtPromptingElicits2022}. This can be achieved under zero-shot settings (prompting the LLM to think step by step)~\cite{kojimaLargeLanguageModels2022} and few-shot settings (prompting the LLM with a few chain-of-thought examples).

\begin{figure*}
\centering
\includegraphics[width=.98\columnwidth]{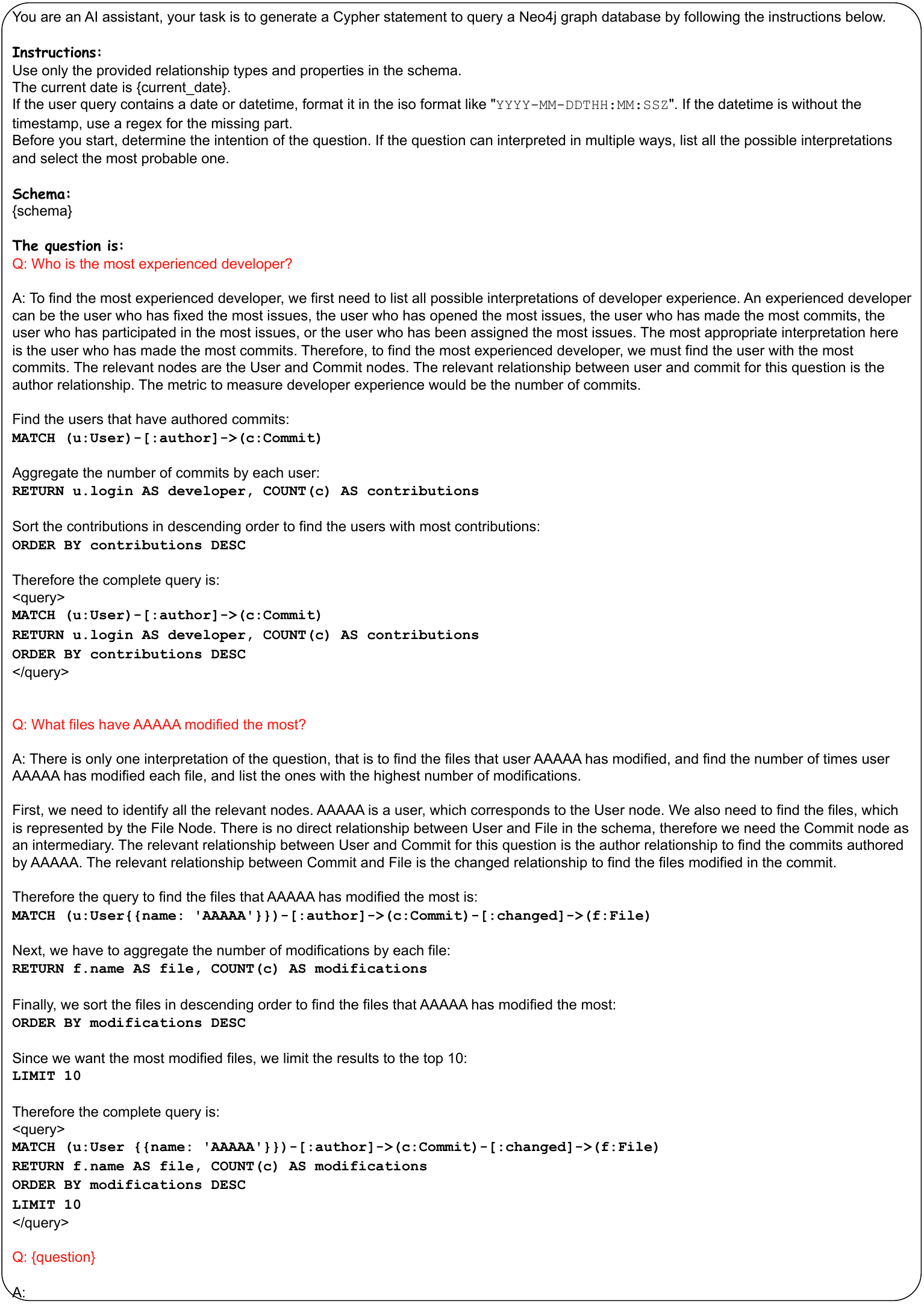}
\caption{Prompt template for the few-shot chain-of-thought. The prompt includes the current date and time, the schema of the knowledge graph, the chain-of-thought examples, and the user's question.}
\label{fig:few-shot_CoT}
\end{figure*}

\smallskip
\noindent\textbf{Approach.} 
To evaluate if chain-of-thought prompting can improve the performance of our approach in answering software repository questions, we first evaluated the approach by including a zero-shot chain-of-thought instruction, that is: \textit{``Let's think step by step''}~\cite{kojimaLargeLanguageModels2022} to the prompt template in Figure~\ref{fig:Cypher_prompt}. This did not improve the results presented in RQ1 (See Appendix~\ref{appendix:b} for the results of the zero-shot chain-of-thought across the selected projects and difficulty levels).  We experimented with few-shot chain-of-thought by incorporating examples into the prompt provided to the Query Generator. Similar to~\citet {weiChainofThoughtPromptingElicits2022}, we adopted the format that begins with the input question, the step-by-step reasoning process, and the final output in our few-shot chain-of-thought prompting. This format aims to guide the LLM in generating intermediate reasoning steps before arriving at the final answer. We constructed the chain-of-thought prompt as presented in Figure~\ref{fig:few-shot_CoT}. The prompt consists of two examples, consisting of difficulty level 2 and 3 questions. By providing these examples, we intended to show the LLM how to generate reasoning paths that can help in constructing correct queries. It is important to note that the questions used in the chain-of-thought examples are not part of our evaluation questions.

For the evaluation, we first measure the gains in execution accuracy made by incorporating chain-of-thought and then benchmark against the baselines described in Section~\ref{sec:baseline_setup}. To evaluate the gains, we use the same set of 20 questions described in Section~\ref{sec:eval_setup} and used in RQ1, executing the experiments five times for each question to account for the stochastic nature of the LLM’s generation~\cite{edgeLocalGlobalGraph2024}. A question is considered correctly answered if the correct response is generated most of the time; otherwise, it is marked as incorrect.

For the baseline comparison, we use the 150 evaluation questions described in Section~\ref{sec:eval_questions} and run each baseline five times, as done previously. The baselines include the intent-based MSRBot and the GPT-4o model with web search capabilities. MSRBot produces identical answers across runs due to its deterministic intent-matching process, while GPT-4o-search-preview is run with the default parameters through the OpenAI API across iterations to ensure a fair comparison.

\smallskip
\noindent\textbf{Results.} 
Table~\ref{tab:repo_comp_rq3} compares the accuracy of the few-shot chain-of-thought approach across the selected project. The accuracy across the projects ranged between 80\% and 90\%, with an average of 84\%. The results improved across all projects to the result in RQ1 (see Table~\ref{tab:repo_comp_rq1}). Table~\ref{tab:level_comp_rq3} also presents the performance based on the difficulty level of the questions. The accuracy improved across all levels with the application of few-shot chain-of-thought prompting compared to without the few-shot chain-of-thought prompt (see Table~\ref{tab:level_comp_rq1}). For level 1 questions, the accuracy increased to 85\% from the previous 80\% in RQ1. For level 2 questions, the accuracy improved to 82\% from 65\% in RQ1. For the level 3 questions, which were previously challenging for our approach, the accuracy significantly increased to 90\% from 50\% in RQ1, indicating a substantial improvement in handling complex queries. Comparing the results with those from RQ1, the overall accuracy improved from 65\% to 84\%, signifying an improvement of 19\% percentage points compared to the results in RQ1.

The improvement in the results, especially in the level 3 questions, suggests that few-shot chain-of-thought prompting effectively mitigated the faulty reasoning limitation identified in RQ2. By providing reasoning steps, our approach could better navigate the complex relationships within the knowledge graph. For example, when asked to \textit{``Determine the percentage of the fixing commits that introduced bugs on July 2023?''}, the few-shot chain-of-thought prompting approach, correctly identified the required relationships between the commits that fixed an issue and, at the same time, introduced another issue to formulate an accurate Cypher query.
However, some limitations persisted. Our approach still encountered difficulties with certain questions, leading to incorrect answers or unanswered questions. In level 1, for instance, the accuracy did not improve as substantially as in level 3, indicating that while chain-of-thought prompting aids in complex reasoning, it may not fully address all types of reasoning errors.

The results with the few-shot chain-of-thought prompting align with prior studies that have shown the effectiveness of chain-of-thought prompting in improving the reasoning abilities of LLMs~\cite{kojimaLargeLanguageModels2022, weiChainofThoughtPromptingElicits2022}. By augmenting the LLM with structured reasoning examples, we facilitated better logical processing, leading to more accurate responses. Chain-of-thought prompting has a positive impact on our approach's performance in answering the questions. It effectively reduces poor reasoning and improves the model's ability to handle complex queries involving multiple relationships.

\noindent\textbf{Comparison with baselines.} 
We further compared our few-shot chain-of-thought approach with the intent-based MSRBot and the GPT-4o model with web search, which serve as baselines for this task (see Section~\ref{sec:baseline_setup}). Table~\ref{tab:baseline_repo_comp_rq3} summarises the accuracy of the three methods across the five projects. Our approach achieved an average accuracy of 0.82 across 750 evaluation questions, outperforming MSRBot (0.70) and GPT-4o-search-preview (0.19). The GPT-4o-search baseline struggled to answer the repository-related questions, corroborating prior observations that retrieval-augmented LLMs often fail, although this approach outperformed the prior study~\cite{abeduLLMBasedChatbotsMining2024}. Table~\ref{tab:baseline_level_comp_rq3} reports the execution accuracy by question difficulty. Our approach consistently outperformed MSRBot on level-1 and level-3 questions, achieving 0.87 accuracy on the most complex level-3 questions compared to 0.60 for MSRBot. For level-2 questions, MSRBot slightly outperformed our approach. GPT-4o-search performed poorly across all levels, reinforcing that naïve web-enabled LLMs are not yet suitable for this task. Overall, few-shot chain-of-thought prompting yields higher accuracy than the baselines on most question types.

\begin{figure}
\centering
\includegraphics[width=.9\columnwidth]{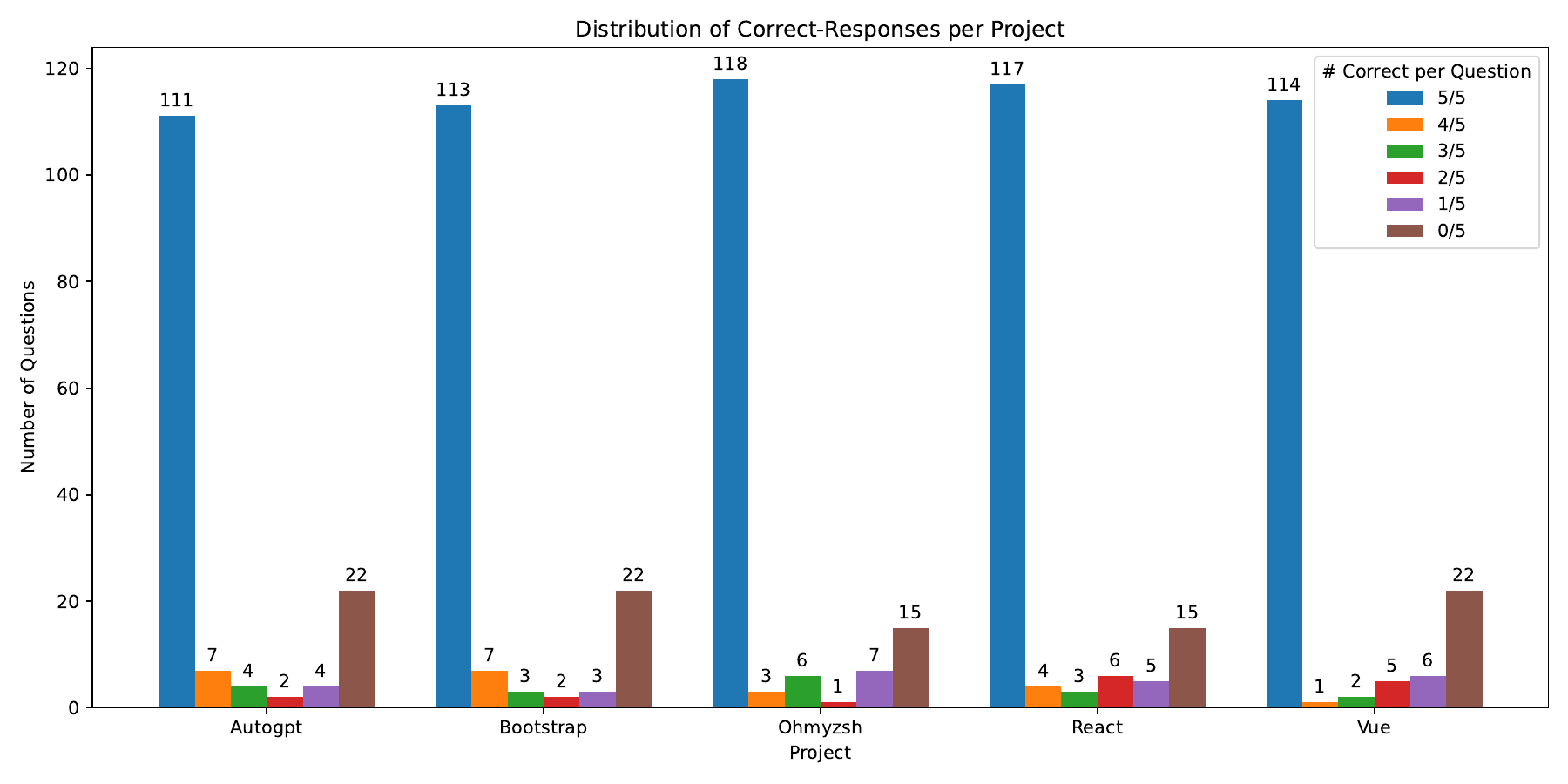}
\caption{Distribution of correct responses across multiple runs for each project. The blue bar represents when all 5 answers are correct, the orange bar when 4 of the 5 answers are correct, the green bar when 3 of 5 are correct, the red bar when 2 of the 5 are correct, the purple bar when 1 of the 5 is correct and the brown bar when all 5 answers are incorrect.}
\label{fig:ratio_correct_responses}
\end{figure}

\noindent\textbf{Consistency of responses.}
Beyond execution accuracy, we examined the \emph{consistency} of responses across multiple runs of the same question for our approach. Specifically, we counted the number of times the correct answer was generated in the five runs for each question. The distribution in Figure~\ref{fig:ratio_correct_responses} shows this for all five projects. Overall, 76.4\% of questions were answered correctly each time the question was repeated (\textit{5/5 correct runs}). This increases to 81.7\% under our majority-vote criterion (i.e., \textit{3/5}, \textit{4/5}, or \textit{5/5} correct runs) in our evaluation.
This shows that our approach is not only accurate on average but also \emph{reliably} produces the correct answer most of the time. This stability is an important quality for practical deployment, as it reduces variability in responses for the user.

\smallskip
\begin{table*}
\caption{Comparison of the accuracy of the few-shot chain-of-thought approach across the selected projects}
\label{tab:repo_comp_rq3}
\centering
\begin{tabular}{lcccc}
\toprule
\textbf{Project} & \textbf{Questions} & \textbf{Answered} & \textbf{Correct} & \textbf{Accuracy} \\
\midrule
\rowcolor{lightgray} AutoGPT    & 20 & 20 & 18 & 0.90  \\
Bootstrap  & 20 & 20 & 16 & 0.80  \\
\rowcolor{lightgray} Ohmyzsh    & 20 & 18 & 16 & 0.80  \\
React      & 20 & 19 & 18 & 0.90  \\
\rowcolor{lightgray} Vue        & 20 & 18 & 16 & 0.80  \\
\midrule
\textbf{Overall}      & \textbf{100} & \textbf{95} & \textbf{84} & \textbf{0.84} \\
\bottomrule
\end{tabular}
\end{table*}

\begin{table*}
\caption{Comparison of the accuracy of the few-shot chain-of-thought approach based on the difficulty level of the questions}
\label{tab:level_comp_rq3}
\centering
\begin{tabular}{lcccc}
\toprule
\textbf{Level} & \textbf{Questions} & \textbf{Answered} & \textbf{Correct} & \textbf{Accuracy} \\
\midrule
\rowcolor{lightgray} 1      & 20  & 18  & 17  & 0.85 \\
2      & 60  & 57  & 49  & 0.82 \\
\rowcolor{lightgray} 3      & 20  & 20  & 18  & 0.90  \\
\midrule
\textbf{Overall}  & \textbf{100} & \textbf{95}  & \textbf{84}  & \textbf{0.84} \\
\bottomrule
\end{tabular}
\end{table*}

\begin{table*}[htbp]
\caption{Comparison with baselines across the selected projects}
\label{tab:baseline_repo_comp_rq3}
\centering
\resizebox{\textwidth}{!}{%
\begin{tabular}{l c c c c c c c}
\toprule
\textbf{Project} & \textbf{Questions} & \multicolumn{3}{c}{\textbf{Correct}} & \multicolumn{3}{c}{\textbf{Accuracy}} \\
\cmidrule(lr){3-5} \cmidrule(lr){6-8}
 &  & \textbf{Our Approach} & \textbf{MSRBot} & \makecell{\textbf{GPT-4o-}\\\textbf{Search-Preview}} 
 & \textbf{Our Approach} & \textbf{MSRBot} & \makecell{\textbf{GPT-4o-}\\\textbf{Search-Preview}} \\
\midrule
\rowcolor{lightgray} AutoGPT    & 150 & 122 & 114 & 26 & 0.81 & 0.76 & 0.17 \\
Bootstrap  & 150 & 123 & 90 & 36 & 0.82 & 0.60 & 0.24 \\
\rowcolor{lightgray} Ohmyzsh    & 150 & 127 & 90 & 32 & 0.85 & 0.60 & 0.21 \\
React      & 150 & 124 & 110 & 24 & 0.83 & 0.73 & 0.16 \\
\rowcolor{lightgray} Vue        & 150 & 117 & 123 & 23 & 0.78 & 0.82 & 0.15 \\
\midrule
\textbf{Overall} & \textbf{750} & \textbf{613} & \textbf{527} & \textbf{141} & \textbf{0.82} & \textbf{0.70} & \textbf{0.19} \\
\bottomrule
\end{tabular}%
}
\end{table*}

\begin{table*}[htbp]
\caption{Comparison with baselines based on the difficulty level of the questions}
\label{tab:baseline_level_comp_rq3}
\centering
\resizebox{\textwidth}{!}{%
\begin{tabular}{l c c c c c c c}
\toprule
\textbf{Level} & \textbf{Questions} & \multicolumn{3}{c}{\textbf{Correct}} & \multicolumn{3}{c}{\textbf{Accuracy}} \\
\cmidrule(lr){3-5} \cmidrule(lr){6-8}
 &  & \textbf{Our Approach} & \textbf{MSRBot} & \makecell{\textbf{GPT-4o-}\\\textbf{Search-Preview}} 
 & \textbf{Our Approach} & \textbf{MSRBot} & \makecell{\textbf{GPT-4o-}\\\textbf{Search-Preview}} \\
\midrule
\rowcolor{lightgray} 1      & 200 & 174 & 123 & 101 & 0.87 & 0.62 & 0.51 \\
2      & 395 & 304 & 311 & 30  & 0.77 & 0.79 & 0.08 \\
\rowcolor{lightgray} 3      & 155 & 135 & 93  & 10  & 0.87 & 0.60 & 0.06 \\
\midrule
\textbf{Overall}  & \textbf{750} & \textbf{613} & \textbf{527} & \textbf{141} & \textbf{0.82} & \textbf{0.70} & \textbf{0.19} \\
\bottomrule
\end{tabular}%
}
\end{table*}

\bigskip
\begin{tcolorbox}
    \paragraph{\emph{\textbf{RQ3 Summary:}}} With few-shot chain-of-thought, the execution accuracy increases from 0.65 to 0.84 overall, and, for complex questions requiring two or more relationships, from 0.50 to 0.90. This implies that chain-of-thought can help answer complex questions requiring multiple relationships. The approach also outperformed the baselines, MSRBot (0.70) and GPT-4o-search-preview (0.19).
\end{tcolorbox}
\medskip

\subsection{RQ4: \rqiv}
\label{rq4}

\noindent\textbf{Motivation.} 
Building on the accuracy gains observed in RQ3, this research question examines whether these improvements translate into practical value for developers. Specifically, we examine whether users find our approach helpful in answering repository-related questions. To this end, we analyze task accuracy, completion time, and qualitative feedback, as these jointly capture both the efficiency gains and the subjective value users derive from the chatbot. This evaluation is critical for validating the approach from the end-user perspective, complementing quantitative performance metrics with insights into perceived utility.

\noindent\textbf{Approach.} 
To evaluate the usefulness of our approach with actual users, we conducted an online user study with 20 participants to assess the perceived usefulness of our approach. The study compared task performance when using our chatbot implementation of the approach against manual task execution. Participants first completed a set of repository-related tasks manually, then repeated similar tasks using the chatbot. We measured task accuracy and time taken, and collected qualitative feedback through a post-task questionnaire (see Section~\ref{sec:survey_design}).

Table~\ref{tab:participant_demographics} shows the breakdown of the demographics of the study participants. Participants comprised 65.0\% students and 35.0\% industry practitioners, with the majority (55.0\%) having over five years of Git experience. Most participants use Git daily (55.0\%).
Although the majority of our participants are students, this does not influence our findings for these reasons: most of the students (53.8\% of student participants) have more than five years of experience using Git, 46.2\% of the student population uses Git daily, and 53.2\% of the student population uses Git weekly. Also, prior studies have shown that students with experience are a good proxy for industry experts, especially when the study is focused on emerging technology~\cite{salman_are_2015}.

\noindent\textbf{Results.}
Each of the 20 participants was tasked with completing 5 repository-related tasks manually and then performing them using the chatbot. The task completion rate and the correctness of the final answers for each participant are presented in Appendix~\ref{appendix:g}. The baseline (manual) approach led to the participants completing 69 out of the 100 tasks and providing a final answer, whereas, when using the chatbot, participants completed 91 tasks. Only three participants completed all five tasks manually, but 15 participants did so when using the chatbot, indicating the chatbot helped users finish more of their work.

Among the tasks that participants completed when working manually, they provided correct answers for 36\% of the tasks, compared with 84\% when using the chatbot. On average, participants solved approximately two tasks correctly without the assistance of our chatbot and more than four tasks correctly using the chatbot. No participant performed worse with the chatbot, and many improved from zero or one correct answer to four or five. These results indicate substantial gains in correctly obtaining information from the repository when using our approach.

\begin{figure}
\centering
\includegraphics[width=.6\columnwidth]{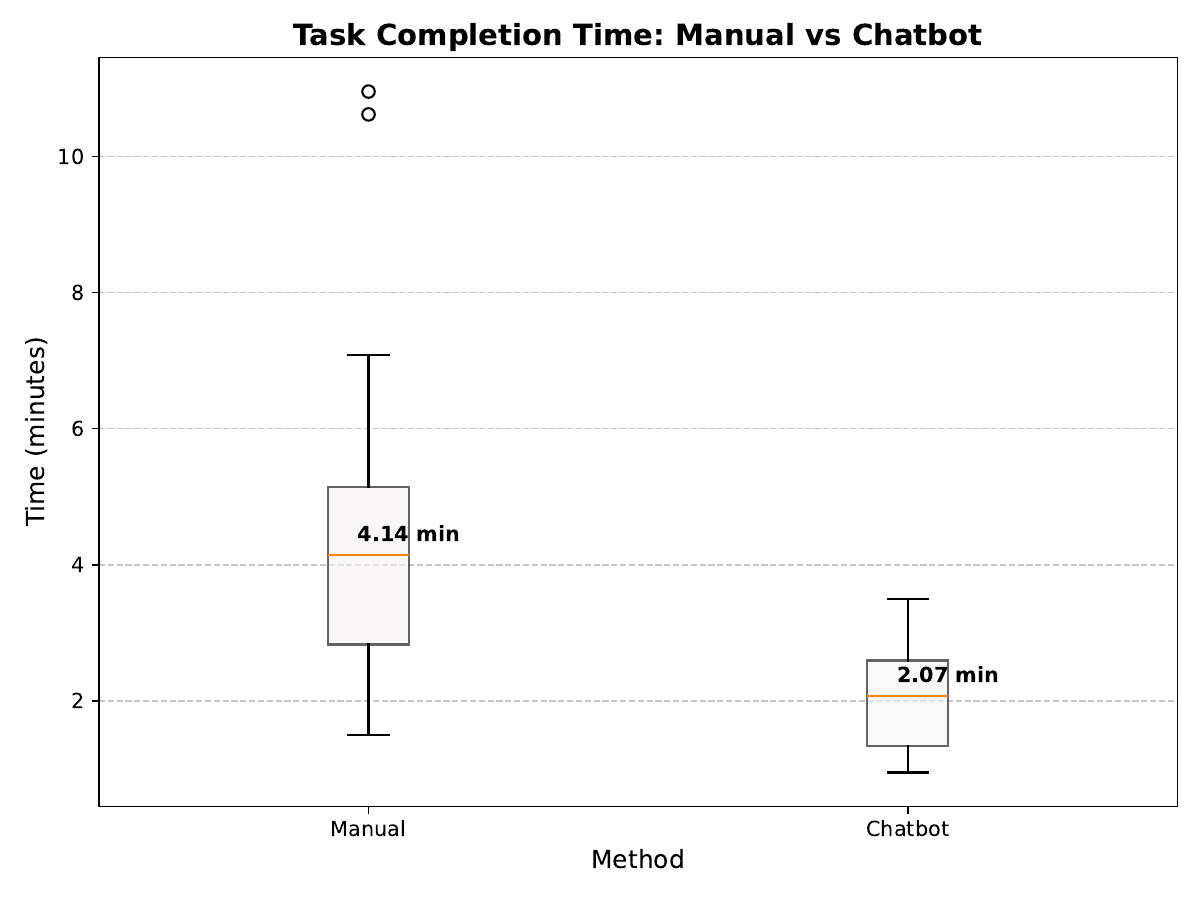}
\caption{Boxplot of task completion times: participants consistently completed tasks faster with the chatbot than with manual methods.}
\label{fig:time_box_plot}
\end{figure}

Efficiency gains were also substantial. Across all five tasks, participants spent a median of 20.7~minutes when working manually, whereas using the chatbot required a median time of 10.3~minutes. Participants typically saved about 12~minutes per session, halving their effort. On a per-task basis, when using manual methods, participants took, on average, 4.5~minutes per answer, while using the chatbot reduced this to just over 2~minutes. Note that the 4.5-minute manual average reflects the 6-minute cap per task and is not an unconstrained effort time.
Figure~\ref{fig:time_box_plot} shows the time participants spent on a task without using the chatbot and when using the chatbot.

The post-survey responses further underscore the chatbot’s perceived usefulness. Participants rated their agreement on a five-point scale and strongly agreed that the chatbot understood their questions (65\% gave it the highest rating of 5), returned correct answers (50\%), made the tasks easier (85\%), and saved them significant time (80\%). A majority also said they preferred using the chatbot over their usual methods (55\% strongly agree) and would recommend it to others (60\% strongly agree), with the remainder generally agreeing. The open-ended comments show a positive sentiment, with users calling the chatbot ``amazing,'' ``very helpful and easy to use,'' and noting that it solved problems they could not handle manually.

Overall, the results show that the chatbot not only boosts completion and accuracy but also substantially reduces effort. These benefits are clearly appreciated by the intended users.

\begin{table}[h]
\centering
\caption{Demographics of Participants in the Survey}
\label{tab:participant_demographics}
\begin{tabular}{llcc}
\hline
\textbf{Category} & \textbf{Experience} & \textbf{Frequency} & \textbf{\%} \\
\hline
Background & Student & 13 & 65 \\
           & Industry Practitioner & 7 & 35 \\
\hline
Years of Experience with Git Tools & 5+ years & 11 & 55 \\
                                    & 3--5 years & 4 & 20 \\
                                    & 1--2 years & 5 & 25 \\
\hline
Frequency of Using Git Tools & Daily & 11 & 55 \\
                              & Weekly & 9 & 45 \\
\hline
\multicolumn{2}{l}{\textbf{Total Participants}} & \textbf{20} & \\
\hline
\end{tabular}
\end{table}

\bigskip
\begin{tcolorbox}
    \paragraph{\emph{\textbf{RQ4 Summary:}}} Users perceived the chatbot as both accurate and efficient for answering their repository-related questions; it doubled the number of correct answers and saved time compared to other methods. Overall, participants found it helpful and preferable to other methods.
\end{tcolorbox}
\medskip

\section{Discussions}
\label{sec:discussions}
In this study, we explored the synergy between large language models (LLMs) and knowledge graphs to enhance the accuracy of software engineering chatbots in answering software repository-related questions. Our findings demonstrate that synergizing LLMs and knowledge graphs can effectively improve the performance of LLM-based chatbots in answering software repositories-related questions. 
Beyond accuracy, our user study in RQ4 shows that these technical gains translate into practical value: participants completed more tasks, answered more correctly, did so in less time when using the chatbot, and reported high perceived usefulness.

We evaluated our approach on five popular open-source GitHub projects, with an accuracy range of 60\%–75\% in RQ1 and 80\%–90\% in RQ3. The difference in the accuracy of the approach on the projects is due to the non-deterministic nature of the LLM rather than the information in the projects. Our approach relies on the schema of the knowledge graph (see Figure~\ref{fig:knowledge_graph}), which is the same for all evaluated projects, to generate a query for retrieving the relevant data. Thus, if the LLM generated the same Cypher query for each question each time the question was executed, the accuracy would be constant across the repositories. This shows that the approach is generalizable across projects.
Supplementing these findings, the baseline comparisons in RQ3 indicate that our method outperforms both MSRBot and the GPT-4o-search-preview model across projects and difficulty levels, reinforcing the benefit of combining knowledge and LLMs with CoT reasoning for repository Q\&A.

We identified faulty reasoning as the prevalent challenge facing the approach, accounting for four of the limitations. This highlights the LLM's challenges in accurately interpreting complex relationships and reasoning over the knowledge graph. In addressing the reasoning limitations, we introduced few-shot chain-of-thought prompting. This technique significantly improved the chatbot's performance by 19\% percentage points compared to the initial accuracy.
In RQ4, this improvement aligned with observed usability gains: participants roughly halved their median task time with the chatbot while increasing correctness, suggesting that better reasoning support helps understand the user's question and subsequently provide a correct answer.

In the evaluation, one aspect we examined was the handling of ambiguous user questions. In RQ3, we instructed the LLM to list all possible interpretations of a question if it detects any ambiguity in the question. The LLM effectively recognized ambiguous questions. For instance, the question \textit{``Which developer has the most number of bugs yet to be fixed?''}, the LLM listed possible interpretations which included \textit{``Developers who have been assigned to these issues''}, \textit{``Developers who have created these issues''}, and  \textit{``Developers who have participated in these issues''} and then selected \textit{``Developers who have been assigned to these issues''} which is the right interpretation. Nonetheless, there were instances where, after listing probable interpretations, it selected the wrong interpretation, leading to an incorrect answer.

Also, the LLM sometimes provided reasonable interpretations of some of the ambiguities, which may lead to correct responses in other contexts. For example, in the question \textit{``Determine the developers that fixed the most bugs in bootstrap-grid.scss?''}. The LLM interpreted it as getting the users that authored commits that modified the file \texttt{bootstrap-grid.scss?} and also fixed bugs. This logic would have been true if the bug was in the file \texttt{bootstrap-grid.scss}. However, if the bug fixing changes are not in the \texttt{bootstrap-grid.scss} file but in a different file, the logic becomes incorrect. The right logic for this question should be getting the users that authored commits which fixed issues that impacted the \texttt{bootstrap-grid.scss} file. In the evaluation in our RQ3, we considered these scenarios as incorrectly answered questions. However, if we had considered such scenarios in our evaluation as correct, the accuracy in RQ3 increased from an average of 84\% to 94\% (See Appendix~\ref{appendix:c} for the comparison of the accuracy of our approach using chain-of-thought prompt across different projects and difficulty levels, when the logic of the LLM is deemed as sound due to ambiguity in the question).

Our approach achieved results beyond the evaluation questions in Section~\ref{sec:eval_setup}. Our approach was able to answer ad-hoc queries that were not part of the original test set. For example, using lines of code as a metric for productivity~\cite{maxwellSoftwareDevelopmentProductivity1996,chengWhatImprovesDeveloper2022}, we asked the question \textit{``Who added the most lines of code in December 2023''} on the Vue project, it responded correctly by going through the right reasoning (See Appendix~\ref{appendix:d}). This highlights the promising potential of using LLMs with knowledge graphs to transform software engineering chatbots, making them more capable of handling a wide range of user queries.
Together with the user study’s time savings and preference ratings, these observations suggest that the approach is not only accurate but also practically useful for everyday repository inquiries.

Another limitation of the generated Cypher query is it uses exact matching when querying the knowledge graph instead of pattern matching. For instance, \textit{``Give me all the commits for vnode.js file?''} returns all the commits that modified the vnode.js file however, \textit{``Give me all the commits for vnode file?''} the response is \textit{``I don't know''} because vnode does not match any filename in the files.

\subsection{Implication for Practitioners.}
The findings of our study have implications for software developers, project managers, and other stakeholders involved in software development. Augmenting chatbots with LLMs and knowledge graphs can significantly enhance accessibility to repository information, making it easier for non-technical team members to retrieve project information without requiring technical expertise. This accessibility can facilitate better collaboration, informed decision-making, and increased efficiency within development teams.

Practitioners like chatbot developers should consider implementing interactive features in chatbots, allowing the chatbot to ask follow-up questions that lead to better understanding and more accurate responses, ultimately improving user satisfaction. This will help improve the chatbot's responses when there is ambiguity. The identified limitations, such as the exact matching instead of the pattern matching in the generated Cypher queries, highlight the need for chatbot developers to include robust error handling and validation in chatbot systems. 

\subsection{Implication for Researchers.}
The findings of this study open avenues for further investigation. Using chain-of-thought improved the accuracy of the reasoning ability of the LLM in our approach. Nonetheless, there are other proposed approaches for enhancing the reasoning ability of LLMs~\cite{kalyanpurLLMARCEnhancingLLMs2024,zhangCanLLMGraph2024,havrillaGLoReWhenWhere2024}. Researchers can build upon this work by investigating other reasoning techniques, integrating symbolic reasoning with neural networks, or exploring alternative prompting strategies to improve the reasoning in LLMs for software engineering chatbots.

The handling of ambiguous queries presents another area for research. Researchers should explore methods for quantifying and reducing ambiguity in user queries. They can focus on developing models that can manage ambiguity by generating multiple interpretations with corresponding confidence levels.

\section{Threats to Validity}
\label{sec:threats}
In this section, we discuss threats to the validity of our study and the measures taken to mitigate them. We consider threats to construct validity, internal validity, and external validity.

\subsection{Construct Validity}
Construct validity pertains to the extent to which our evaluation measures accurately reflect the theoretical constructs they are intended to assess. A threat to construct validity is the dependency on the knowledge graph schema. The LLM's ability to generate a correct Cypher query is dependent on its understanding of the schema of the knowledge graph. Any discrepancies or ambiguities in the schema can lead to incorrect Cypher query generation. If the knowledge graph schema does not accurately represent the repository data, the LLM may produce queries that do not retrieve the intended information. We mitigated this by providing explanations of the meaning of the relationships between the entities in the knowledge graph.

\subsection{Internal Validity}
Internal validity refers to the extent to which the observed effects can be attributed to the variables under investigation rather than other factors. A threat to internal validity in our study is the stochastic nature of LLM outputs. Despite setting the temperature parameter to zero to reduce randomness, inherent variability in the LLM's responses could influence the results. Correct answers might occasionally occur by chance rather than due to the effectiveness of our approach.

To address this, we executed the experiment for each question five times and used a majority-vote criterion to determine correctness. A question was considered correctly answered if the chatbot provided the correct response at least 50\% of the time (three out of the five attempts). This approach aimed to mitigate situations where a question was correctly answered by chance or otherwise. 

Another threat to internal validity is related to the data dependencies in our approach. Specifically, if the bug ID or issue number is not specified in the commit log of the fixing commit, our approach will not identify it as such. The identification of bug-fixing commits in our approach relies on these references to link commits to the issues accurately. Missing or inconsistent references can affect the completeness of information in the knowledge graph. However, this approach of identifying the fixing commits for issues follows procedures presented in prior studies~\cite{dacostaFrameworkEvaluatingResults2017}

\subsection{External Validity}
External validity concerns the generalizability of our findings beyond this study. A first threat is generalizing beyond the evaluation questions. We evaluated on 150 questions covering 10 intents (Section~\ref{sec:eval_setup}). Although these questions cover different difficulty levels and intents, they may not encompass the full diversity of questions that users might pose in real-world scenarios. To mitigate this, we ran a task-based user study with 20 participants (Section~\ref{sec:survey_design}) where users phrased tasks in their own words and worked both manually and with our chatbot. We also posed ad hoc questions outside the benchmark. These observations suggest some generalizability, but different question types could yield different results.

A second threat is the scope of our knowledge graph. The current schema models four entity types (Users, Commits, Issues, and Files), so questions that depend on non-modelled entities (e.g., pull requests, releases/tags, or CI/CD build events) cannot be answered. Although these four entities are critical to many repository-related questions, other entities could be added to the knowledge graph to answer more questions.

\section{Conclusion}
\label{sec:conclusion}
In this study, we investigated the synergy between large language models (LLMs) and knowledge graphs to improve the accuracy of software engineering chatbots in answering software repository-related questions. Our approach aimed to accurately answer natural language questions by generating Cypher queries to retrieve relevant repository data from the knowledge graph. Then use the retrieved information as context to generate a natural language response, making repository information accessible to both technical and non-technical stakeholders. We empirically evaluated our approach using five popular open-source GitHub repositories and a set of 20 questions curated from~\citet{abdellatifMSRBotUsingBots2020} and categorized into three levels of difficulties. The findings demonstrated that LLMs, specifically the GPT-4o model, can answer repository-related questions by generating Cypher queries to retrieve accurate data from the knowledge graph. The initial accuracy of 65\% achieved by our approach highlighted the potential limitation of synergizing LLMs with knowledge graphs. We manually investigated the instances where the approach failed to generate an accurate response and identified the faulty reasoning by the LLM as the predominant factor (80.5\%) affecting the approach. We conducted further empirical evaluation if using few-shot chain-of-thought prompting can improve the accuracy. This technique significantly enhanced the reasoning ability of the LLM in our approach and improved the overall accuracy from 65\% to 84\%. There was a notable increase in the accuracy of the level 3 questions from 50\% to 90\%, signifying an improvement in the approach to handling complex queries. In addition to these gains, the baseline comparisons showed that combining LLMs with knowledge outperforms the intent-based approach, MSRBot, and GPT-4o-search-preview across projects and difficulty levels. Our findings highlight the integration of LLMs with knowledge graphs as a viable solution for making repository data accessible to both technical and non-technical stakeholders. The user study demonstrated that the technical improvements translate into practical benefits. In the study, the participants completed more tasks correctly and in less time with the chatbot than with their usual methods, and they reported high perceived usefulness. Also, our study highlights the importance of enhancing reasoning capabilities in LLMs.  This opens avenues for further investigation in this direction.
In this study, we focus on Git metadata (Users, Commits, Issues, Files) and do not answer questions that require program analysis. Future work can extend the repository knowledge graph with code entities and program-analysis edges and align the code-level entities with the existing metadata to answer questions about code comprehension.

\bibliographystyle{ACM-Reference-Format}
\bibliography{references}

\clearpage
\appendix
\clearpage
\section{Questions used for the evaluation of the approach}
\label{appendix:a}

\begin{table}[H]
\caption{Questions with parameters and corresponding intents and difficulty levels}
\label{tab:questions}
\centering
\begin{tabularx}{\textwidth}{c >{\raggedright\arraybackslash}X c c}
\toprule
\textbf{\#} & \textbf{Question} & \textbf{Intent} & \textbf{Level} \\
\midrule
\rowcolor{lightgray} 1 & How many commits happened in \texttt{[DATE RANGE]}? & Commits By Date Period & 1 \\
2 & What is the latest commit? & Commits By Date & 1 \\
\rowcolor{lightgray} 3 & Can you tell me the details of the commits between \texttt{[DATE RANGE]}? & Commits By Date Period & 1 \\
4 & Return a commit message on \texttt{[DATE]}? & Commits By Date & 1 \\
\rowcolor{lightgray} 5 & Show me the changes for \texttt{[FILENAME]} file? & File Commits & 2 \\
6 & Give me all the commits for \texttt{[FILENAME]} file? & File Commits & 2 \\
\rowcolor{lightgray} 7 & Determine the developers that had the most unfixed bugs? & Overloaded Dev & 2 \\
8 & Which developer has most number of bugs yet to be fixed? & Overloaded Dev & 2 \\
\rowcolor{lightgray} 9 & Determine the developers that fixed the most bugs in \texttt{[FILENAME]}? & Experienced Dev Fix Bugs & 3 \\
10 & Who did most fixed bugs in \texttt{[FILENAME]}? & Experienced Dev Fix Bugs & 3 \\
\rowcolor{lightgray} 11 & Determine the files that introduce the most bugs? & Buggy Files & 2 \\
12 & What are the most buggy files? & Buggy Files & 2 \\
\rowcolor{lightgray} 13 & What are the buggy commits that happened on \texttt{[DATE]}? & Buggy Commits By Date & 2 \\
14 & What commits were buggy on \texttt{[DATE]}? & Buggy Commits By Date & 2 \\
\rowcolor{lightgray} 15 & Commit(s) that fixed the bug ticket \texttt{[ISSUE ID]}? & Fix Commit & 2 \\
16 & Which commit fixed the bug ticket \texttt{[ISSUE ID]}? & Fix Commit & 2 \\
\rowcolor{lightgray} 17 & Determine the bug(s) that were introduced because of commit hash \texttt{[COMMIT HASH]}? & Buggy Commits & 2 \\
18 & What are the bugs caused by commit \texttt{[COMMIT HASH]}? & Buggy Commits & 2 \\
\rowcolor{lightgray} 19 & Determine the percentage of the fixing commits that introduced bugs on \texttt{[DATE]}? & Buggy Fix Commits & 3 \\
20 & How many fixing commits caused bugs on \texttt{[DATE]}? & Buggy Fix Commits & 3 \\
\bottomrule
\end{tabularx}
\end{table}

\clearpage
\section{Question answered correctly and incorrectly in each project in RQ1}
\label{appendix:e}

\begin{table}[H]
\centering
\caption{Questions answered in each project. \cmark indicates the question was answered correctly and \xmark indicates the question was answered incorrectly or not answered}
\label{tab:answered_rq1}
\begin{tabularx}{\textwidth}{c X c c c c c}
\toprule
\textbf{\#} & \textbf{Question} & \textbf{AutoGPT} & \textbf{Bootstrap} & \textbf{Ohmyzsh} & \textbf{React} & \textbf{Vue} \\
\midrule
\rowcolor{lightgray} 1  & How many commits happened in \texttt{[DATE RANGE]}?                             & \cmark & \cmark & \cmark & \cmark & \cmark \\
2  & What is the latest commit?                                                      & \cmark & \cmark & \cmark & \cmark & \cmark \\
\rowcolor{lightgray} 3  & Can you tell me the details of the commits between \texttt{[DATE RANGE]}?       & \cmark & \cmark & \cmark & \cmark & \xmark \\
4  & Return a commit message on \texttt{[DATE]}?                                     & \cmark & \xmark & \xmark & \cmark & \xmark \\
\rowcolor{lightgray} 5  & Show me the changes for \texttt{[FILENAME]} file?                               & \cmark & \cmark & \cmark & \cmark & \cmark \\
6  & Give me all the commits for \texttt{[FILENAME]} file?                           & \cmark & \cmark & \cmark & \cmark & \cmark \\
\rowcolor{lightgray} 7  & Determine the developers that had the most unfixed bugs?                        & \xmark & \xmark & \xmark & \xmark & \xmark \\
8  & Which developer has most number of bugs yet to be fixed?                        & \xmark & \cmark & \cmark & \xmark & \xmark \\
\rowcolor{lightgray} 9  & Determine the developers that fixed the most bugs in \texttt{[FILENAME]}?       & \cmark & \cmark & \cmark & \cmark & \cmark \\
10 & Who did most fixed bugs in \texttt{[FILENAME]}?                                 & \cmark & \xmark & \cmark & \cmark & \cmark \\
\rowcolor{lightgray} 11 & Determine the files that introduce the most bugs?                               & \cmark & \xmark & \cmark & \xmark & \xmark \\
12 & What are the most buggy files?                                                  & \xmark & \cmark & \cmark & \cmark & \cmark \\
\rowcolor{lightgray} 13 & What are the buggy commits that happened on \texttt{[DATE]}?                    & \cmark & \cmark & \xmark & \xmark & \cmark \\
14 & What commits were buggy on \texttt{[DATE]}?                                     & \cmark & \cmark & \xmark & \xmark & \xmark \\
\rowcolor{lightgray} 15 & Commit(s) that fixed the bug ticket \texttt{[ISSUE ID]}?                        & \cmark & \cmark & \xmark & \cmark & \cmark \\
16 & Which commit fixed the bug ticket \texttt{[ISSUE ID]}?                          & \cmark & \cmark & \cmark & \cmark & \cmark \\
\rowcolor{lightgray} 17 & Determine the bug(s) that were introduced because of commit \texttt{[COMMIT HASH]}? & \cmark & \cmark & \cmark & \cmark & \cmark \\
18 & What are the bugs caused by commit \texttt{[COMMIT HASH]}?                      & \cmark & \xmark & \xmark & \cmark & \cmark \\
\rowcolor{lightgray} 19 & Determine the percentage of the fixing commits that introduced bugs on \texttt{[DATE]}? & \xmark & \xmark & \xmark & \xmark & \xmark \\
20 & How many fixing commits caused bugs on \texttt{[DATE]}?                         & \xmark & \xmark & \xmark & \xmark & \xmark \\
\bottomrule
\end{tabularx}
\end{table}

\clearpage
\section{Question answered correctly and incorrectly in each project in RQ3}
\label{appendix:f}

\begin{table}[H]
\centering
\caption{Questions answered in each project with few-shot chain-of-thought prompting. \cmark indicates the question was answered correctly and \xmark indicates the question was answered incorrectly or not answered}
\label{tab:answered_rq3}
\begin{tabularx}{\textwidth}{c X c c c c c}
\toprule
\textbf{\#} & \textbf{Question} & \textbf{AutoGPT} & \textbf{Bootstrap} & \textbf{Ohmyzsh} & \textbf{React} & \textbf{Vue} \\
\midrule
\rowcolor{lightgray} 1  & How many commits happened in \texttt{[DATE RANGE]}?                             & \cmark & \cmark & \cmark & \cmark & \cmark \\
2  & What is the latest commit?                                                      & \cmark & \cmark & \cmark & \cmark & \cmark \\
\rowcolor{lightgray} 3  & Can you tell me the details of the commits between \texttt{[DATE RANGE]}?       & \cmark & \cmark & \cmark & \cmark & \cmark \\
4  & Return a commit message on \texttt{[DATE]}?                                     & \cmark & \xmark & \xmark & \cmark & \xmark \\
\rowcolor{lightgray} 5  & Show me the changes for \texttt{[FILENAME]} file?                               & \cmark & \cmark & \cmark & \cmark & \cmark \\
6  & Give me all the commits for \texttt{[FILENAME]} file?                           & \cmark & \cmark & \cmark & \cmark & \cmark \\
\rowcolor{lightgray} 7  & Determine the developers that had the most unfixed bugs?                        & \xmark & \xmark & \xmark & \xmark & \xmark \\
8  & Which developer has most number of bugs yet to be fixed?                        & \cmark & \cmark & \cmark & \cmark & \cmark \\
\rowcolor{lightgray} 9  & Determine the developers that fixed the most bugs in \texttt{[FILENAME]}?       & \cmark & \cmark & \xmark & \cmark & \cmark \\
10 & Who did most fixed bugs in \texttt{[FILENAME]}?                                 & \cmark & \xmark & \cmark & \cmark & \cmark \\
\rowcolor{lightgray} 11 & Determine the files that introduce the most bugs?                               & \xmark & \xmark & \cmark & \cmark & \xmark \\
12 & What are the most buggy files?                                                  & \cmark & \cmark & \cmark & \cmark & \cmark \\
\rowcolor{lightgray} 13 & What are the buggy commits that happened on \texttt{[DATE]}?                    & \cmark & \cmark & \cmark & \cmark & \cmark \\
14 & What commits were buggy on \texttt{[DATE]}?                                     & \cmark & \cmark & \xmark & \xmark & \xmark \\
\rowcolor{lightgray} 15 & Commit(s) that fixed the bug ticket \texttt{[ISSUE ID]}?                        & \cmark & \cmark & \cmark & \cmark & \cmark \\
16 & Which commit fixed the bug ticket \texttt{[ISSUE ID]}?                          & \cmark & \cmark & \cmark & \cmark & \cmark \\
\rowcolor{lightgray} 17 & Determine the bug(s) that were introduced because of commit \texttt{[COMMIT HASH]}? & \cmark & \cmark & \cmark & \cmark & \cmark \\
18 & What are the bugs caused by commit \texttt{[COMMIT HASH]}?                      & \cmark & \cmark & \cmark & \cmark & \cmark \\
\rowcolor{lightgray} 19 & Determine the percentage of the fixing commits that introduced bugs on \texttt{[DATE]}? & \cmark & \cmark & \cmark & \cmark & \cmark \\
20 & How many fixing commits caused bugs on \texttt{[DATE]}?                         & \cmark & \cmark & \cmark & \cmark & \cmark \\
\bottomrule
\end{tabularx}
\end{table}

\clearpage
\section{Results of the zero-shot chain-of-thought prompting across the selected projects and difficulty levels}
\label{appendix:b}

\begin{table}[H]
\caption{Comparison of the accuracy of the zero-shot chain-of-thought prompting across the selected projects}
\label{tab:repo_comp_appendix_b}
\centering
\begin{tabular}{lcccc}
\toprule
\textbf{Project} & \textbf{Questions} & \textbf{Answered} & \textbf{Correct} & \textbf{Accuracy} \\
\midrule
\rowcolor{lightgray} AutoGPT   & 20  & 20 & 14 & 0.70 \\
Bootstrap & 20  & 19 & 15 & 0.75 \\
\rowcolor{lightgray} Ohmyzsh   & 20  & 19 & 15 & 0.75 \\
React     & 20  & 20 & 15 & 0.75 \\
\rowcolor{lightgray} Vue       & 20  & 19 & 11 & 0.55 \\
\midrule
\textbf{Overall} & \textbf{100} & \textbf{97} & \textbf{70} & \textbf{0.70} \\
\bottomrule
\end{tabular}
\end{table}

\begin{table}[H]
\caption{Comparison of the accuracy of the zero-shot chain-of-thought prompting based on the difficulty level of the questions}
\label{tab:level_comp_appendix_b}
\centering
\begin{tabular}{lcccc}
\toprule
\textbf{Level} & \textbf{Questions} & \textbf{Answered} & \textbf{Correct} & \textbf{Accuracy} \\
\midrule
\rowcolor{lightgray} 1     & 20  & 20 & 18 & 0.90 \\
2     & 60  & 58 & 40 & 0.67 \\
\rowcolor{lightgray} 3     & 20  & 19 & 12 & 0.60 \\
\midrule
\textbf{Overall} & \textbf{100} & \textbf{97} & \textbf{70} & \textbf{0.70} \\
\bottomrule
\end{tabular}
\end{table}

\section{Additional Results on the selected projects and difficulty levels when the reasoning of the LLM on the ambiguous question are evaluated as correct}
\label{appendix:c}

\begin{table}[H]
\caption{Results by project showing the chain-of-thought responses where the LLM's reasoning was considered correct in ambiguous questions}
\label{tab:repo_comp_appendix_c}
\centering
\begin{tabular}{lcccc}
\toprule
\textbf{Project} & \textbf{Questions} & \textbf{Answered} & \textbf{Correct} & \textbf{Accuracy} \\
\midrule
\rowcolor{lightgray} AutoGPT  & 20 & 20 & 20 & 1.00 \\
Bootstrap  & 20 & 20 & 19 & 0.95 \\
\rowcolor{lightgray} Ohmyzsh    & 20 & 18 & 18 & 0.90 \\
React      & 20 & 19 & 19 & 0.95 \\
\rowcolor{lightgray} Vue        & 20 & 18 & 18 & 0.90 \\
\midrule
\textbf{Overall}      & \textbf{100} & \textbf{95} & \textbf{94} & \textbf{0.94} \\
\bottomrule
\end{tabular}
\end{table}

\begin{table}[H]
\caption{Results by level showing the chain-of-thought responses where the LLM's reasoning was considered correct in ambiguous questions}
\label{tab:level_comp_appendix_c}
\centering
\begin{tabular}{lcccc}
\toprule
\textbf{Level} & \textbf{Questions} & \textbf{Answered} & \textbf{Correct} & \textbf{Accuracy} \\
\midrule
\rowcolor{lightgray} 1      & 20  & 18 & 17 & 0.85 \\
2      & 60  & 57 & 57 & 0.95 \\
\rowcolor{lightgray} 3      & 20  & 20 & 20 & 1.00 \\
\midrule
\textbf{Overall}  & \textbf{100} & \textbf{95}  & \textbf{94}  & \textbf{0.94} \\
\bottomrule
\end{tabular}
\end{table}

\section{Number of Tasks Completed by Each Participant and Their Correct Answers With and Without the Chatbot}
\label{appendix:g}

\begin{table*}[htbp]
\centering
\caption{Task Completion and Correctness per Participant (Manual vs.\ Chatbot).}
\resizebox{\linewidth}{!}{%
\begin{tabular}{l c c c c c}
\toprule
\textbf{Participant} & \textbf{Tasks Assigned} & \textbf{Completed Manually} & \textbf{Completed with Chatbot} & \textbf{Manual Correct} & \textbf{Chatbot Correct} \\
\midrule
\rowcolor{lightgray} Participant 1  & 5 & 0 & 5 & 0 & 5 \\
Participant 2  & 5 & 4 & 5 & 4 & 5 \\
\rowcolor{lightgray} Participant 3  & 5 & 4 & 5 & 3 & 5 \\
Participant 4  & 5 & 5 & 4 & 2 & 4 \\
\rowcolor{lightgray} Participant 5  & 5 & 2 & 5 & 2 & 5 \\
Participant 6  & 5 & 3 & 5 & 2 & 4 \\
\rowcolor{lightgray} Participant 7  & 5 & 5 & 5 & 4 & 5 \\
Participant 8  & 5 & 5 & 5 & 4 & 5 \\
\rowcolor{lightgray} Participant 9  & 5 & 4 & 5 & 3 & 5 \\
Participant 10 & 5 & 4 & 4 & 0 & 3 \\
\rowcolor{lightgray} Participant 11 & 5 & 4 & 4 & 2 & 4 \\
Participant 12 & 5 & 3 & 4 & 1 & 1 \\
\rowcolor{lightgray} Participant 13 & 5 & 4 & 5 & 1 & 5 \\
Participant 14 & 5 & 4 & 5 & 2 & 4 \\
\rowcolor{lightgray} Participant 15 & 5 & 4 & 5 & 2 & 4 \\
Participant 16 & 5 & 4 & 5 & 1 & 5 \\
\rowcolor{lightgray} Participant 17 & 5 & 3 & 5 & 1 & 5 \\
Participant 18 & 5 & 3 & 5 & 1 & 5 \\
\rowcolor{lightgray} Participant 19 & 5 & 0 & 0 & 0 & 0 \\
Participant 20 & 5 & 4 & 5 & 1 & 5 \\
\midrule
\textbf{Total} & \textbf{100} & \textbf{69} & \textbf{91} & \textbf{36} & \textbf{84} \\
\bottomrule
\end{tabular}%
}
\end{table*}

\clearpage
\section{Ad-hoc example demonstration}
\label{appendix:d}

\begin{figure}[H]
\centering
\includegraphics[width=.95\columnwidth]{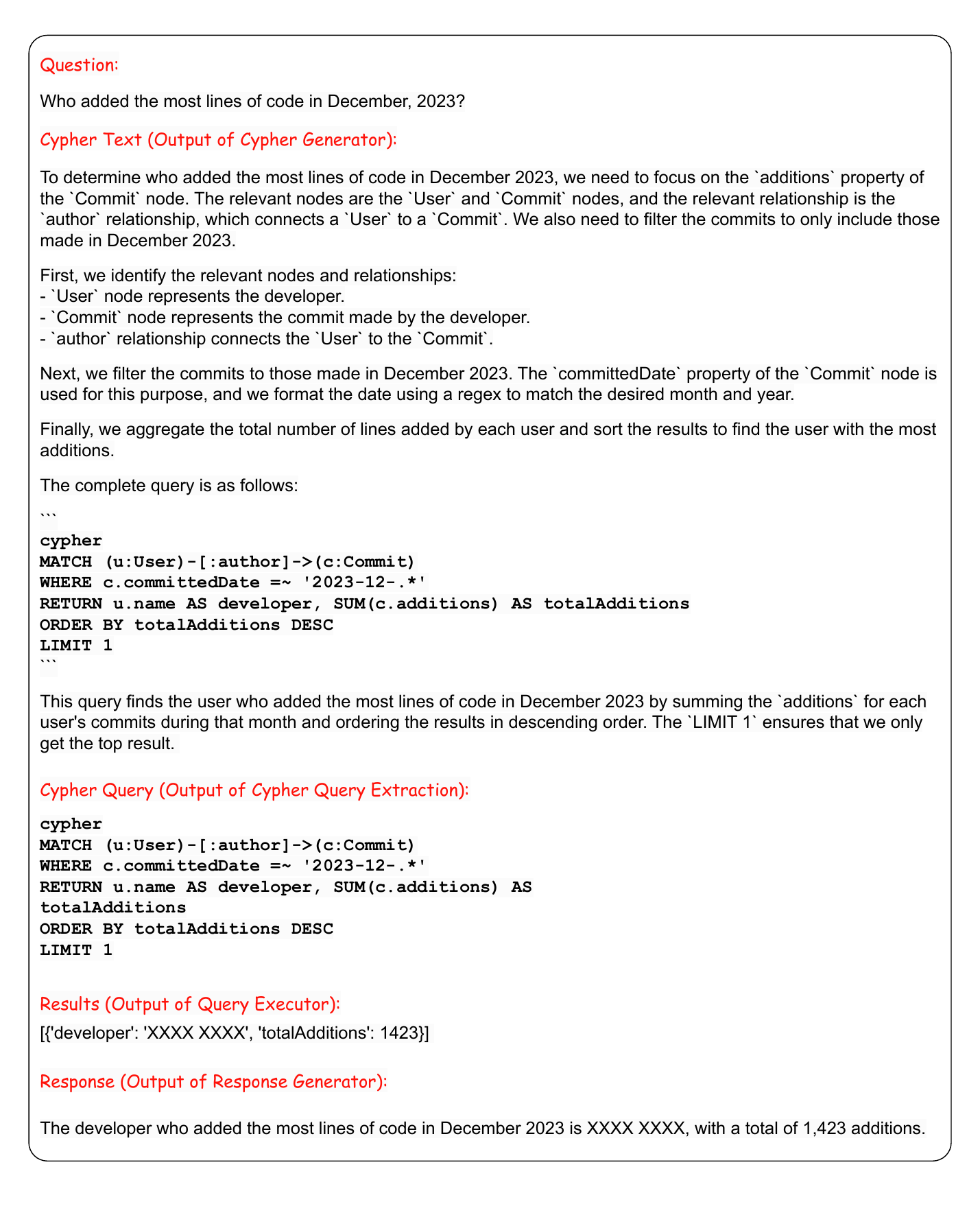}
\caption{Demonstration of the approach answering a question not in the evaluation set. To protect privacy, we have redacted the developer's name}
\end{figure}

\end{document}